\documentclass[useAMS,usenatbib]{mn2e}
\usepackage{times}
\usepackage{color,url,amsmath}
\usepackage{graphicx}
\newcommand{\bmf}[1]{\mbox{\boldmath$#1$}}
\newcommand{\simgt}{\lower.5ex\hbox{$\; \buildrel > \over \sim \;$}}
\newcommand{\simlt}{\lower.5ex\hbox{$\; \buildrel < \over \sim \;$}}

\begin{document}
\title[Weak Lensing Measurement of ACT-CL J0022.2$-$0036]{Subaru
weak-lensing measurement of a $\bf{z = 0.81}$ cluster discovered by the
Atacama Cosmology Telescope Survey}

\author[Miyatake et al.]{Hironao~Miyatake$^{1,2,3}$, Atsushi~J.~Nishizawa$^2$,
Masahiro~Takada$^2$, Rachel~Mandelbaum$^{3,4}$, 
\newauthor 
Sogo~Mineo$^{1,2}$, Hiroaki~Aihara$^{1,2}$, David~N.~Spergel$^{2,3}$, Steven J. Bickerton$^{2,3}$, 
\newauthor
J. Richard Bond$^5$, Amir Hajian$^5$, Matt Hilton$^{6,7}$, Adam D. Hincks$^5$, John P. Hughes$^8$, 
\newauthor
Leopoldo Infante$^9$, Yen-Ting Lin$^{10}$, Robert
H. Lupton$^3$, Tobias A. Marriage$^{11}$, 
\newauthor
Danica Marsden$^{12}$, Felipe Menanteau$^8$, Satoshi
Miyazaki$^{13}$, Kavilan Moodley$^{7}$,
\newauthor
Michael D. Niemack$^{14}$, Masamune Oguri$^{2}$, Paul A. Price$^{3}$, Erik D. Reese$^{15}$, 
\newauthor
Crist\'obal Sif\'on$^{9, 16}$, Edward J. Wollack$^{17}$, Naoki Yasuda$^{2}$\\
$^1$ Department of Physics, The University of Tokyo, Bunkyo, Tokyo 113-0031,
Japan\\
$^2$ Kavli Institute for the Physics and Mathematics of the Universe
(Kavli IPMU, WPI), The University of Tokyo, Kashiwa, Chiba 277-8582, Japan \\
$^3$ Department of Astrophysical Sciences, Princeton University, Princeton, NJ 08544, USA \\
$^4$ Department of Physics, Carnegie Mellon University, Pittsburgh, PA
15213, USA\\
$^5$ Canadian Institute for Theoretical Astrophysics, University of
Toronto, Toronto, ON M5S 3H8, Canada\\
$^6$ School of Physics and Astronomy, University of Nottingham,
University Park, Nottingham, NG7 2RD, UK\\
$^7$ Astrophysics \& Cosmology Research Unit, School of Mathematics, Statistics \& Computer Science, University of KwaZulu-Natal, Durban 4041, South Africa\\
$^8$ Department of Physics and Astronomy, Rutgers, The State University
of New Jersey, Piscataway, NJ 08854, USA\\
$^9$ Departamento de Astronom\'ia y Astrof\'isica, Facultad de F\'isica,
Pontificia Universidad Cat\'olica de Chile, Casilla 306, Santiago 22, Chile\\
$^{10}$ Institute of Astronomy and Astrophysics, Academia Sinica, Taipei, Taiwan\\
$^{11}$ Department of Physics and Astronomy, The Johns Hopkins
University, Baltimore, MD 21218, USA\\
$^{12}$ Department of Physics, University of California, Santa Barbara, CA 93106, USA\\
$^{13}$ National Astronomical Observatory of Japan, Mitaka, Tokyo
181-8588, Japan\\
$^{14}$ NIST Quantum Devices Group, Boulder, CO 80305, USA\\
$^{15}$ Department of Physics and Astronomy, University of Pennsylvania,
209 South 33rd Street, Philadelphia, PA 19104, USA \\
$^{16}$ Leiden Observatory, Leiden University, NL-2300 RA Leiden,
Netherlands \\
$^{17}$ NASA/Goddard Space Flight Center, Greenbelt, MD
20771, USA \\
}
\maketitle

\begin{abstract}
We present a Subaru weak lensing measurement of ACT-CL J0022.2$-$0036,
one of the most luminous, high-redshift ($z=0.81$) Sunyaev-Zel'dovich
(SZ) clusters discovered in the 268 deg$^2$ equatorial region survey of
the Atacama Cosmology Telescope that overlaps with SDSS Stripe 82
field. Ours is the first weak lensing study with Subaru at such high
redshifts.  For the weak lensing analysis using $i'$-band images, we use
a model-fitting (Gauss-Laguerre shapelet) method to measure shapes of
galaxy images, where we fit galaxy images in different exposures
simultaneously to obtain best-fit ellipticities taking into account the
different PSFs in each exposure. We also take into account the
astrometric distortion effect on galaxy images by performing the model
fitting in the world coordinate system. To select background galaxies
behind the cluster at $z=0.81$, we use photometric redshift (photo-$z$)
estimates for every galaxy derived from the co-added images of
multi-passband $Br'i'z'Y$, with PSF matching/homogenization.  After a
photo-$z$ cut for background galaxy selection, we detect the tangential
weak lensing distortion signal with a total signal-to-noise ratio of
about 3.7.  By fitting a Navarro-Frenk-White model to the measured shear
profile, we find the cluster mass to be $M_{200\bar{\rho}_m} =
\left[7.5^{+3.2}_{-2.8}({\rm stat.})^{+1.3}_{-0.6}({\rm
sys.})\right]\times10^{14} M_\odot/h$. The weak lensing-derived mass is
consistent with previous mass estimates based on the SZ observation,
with assumptions of hydrostatic equilibrium and virial theorem, as well
as with scaling relations between SZ signal and mass derived from weak
lensing, X-ray, and velocity dispersion, within the measurement errors.
We also show that the existence of ACT-CL J0022.2$-$0036 at $z=0.81$ is
consistent with the cluster abundance prediction of the
$\Lambda$-dominated cold dark matter structure formation model.  We thus
demonstrate the capability of Subaru-type ground-based images for
studying weak lensing of high-redshift clusters.
\end{abstract}

\section{Introduction}
\label{sec:intro}

Clusters of galaxies are the most massive gravitationally-bound objects
in the Universe, and therefore are very sensitive to cosmological
parameters, including the dark energy equation of state \citep[][and
references therein]{Kitayama:1997,Vikhlinin:2009}.  The growth of cosmic
structures in the Universe is regulated by a competition between
gravitational attraction and cosmic expansion. Hence, if the evolution
of the cluster mass function can be measured robustly, the influence of
dark energy on the growth of structure, and thus the nature of dark
energy, can be extracted. Furthermore, since dark matter plays an
essential role in the formation and evolution of clusters, the mass
distribution in cluster regions contains a wealth of information on the
nature of dark matter
\citep[e.g.,][]{Broadhurst:2005,Okabe:2010,Oguri:2012}.

The Sunyaev-Zel'dovich (SZ) effect, in which photons of the cosmic
microwave background (CMB) scatter off electrons of the hot intracluster
medium, is a powerful way of finding massive clusters, especially at
high redshift (\citealt{Zeldovich:1969}; \citealt{Sunyaev:1972}; also
see \citealt{Carlstrom:2002} for a thorough review), for several
reasons. First, the SZ effect has a unique frequency dependence: below
218GHz, it appears as a decrement (or cold spot) in the CMB temperature
map, while at higher frequencies it appears as an increment (hot
spot). Second, unlike optical and X-ray observations, the SZ effect does
not suffer from the cosmological surface brightness-dimming effect;
thus, it is independent of redshift, offering a unique way of detecting
all clusters above some mass limit irrespective of their
redshifts. Currently there are several ongoing arcminute-resolution,
high-sensitivity CMB experiments, such as the Atacama Cosmology
Telescope (ACT; \citealt{Swetz:2011}) and the South Pole Telescope (SPT;
\citealt{Carlstrom:2011}). These SZ surveys are demonstrating the power
of SZ surveys for finding clusters \citep{Marriage:2011}, and have
already shown that the SZ-detected clusters can be used to constrain
cosmology \citep{Vanderlinde:2010, Sehgal:2011,Reichardt:2012}.

However, the SZ effect itself does not necessarily provide robust mass
estimates of high-redshift clusters, because of several assumptions that
may not be valid, such as dynamical and hydrostatic equilibrium, or the
cluster mass-scaling relation inferred from low-redshift clusters.  The
relationship between cluster observables and mass is of critical
importance for cluster-based cosmology, so it is critical to establish a
well-calibrated scaling relation in order to robustly use SZ-detected
clusters for cosmology.  Gravitational weak lensing (WL), the shape
distortion of background galaxies due to the mass in clusters, is a
well-known tool for unveiling the distribution of matter in clusters,
regardless of the dynamical state \citep[see][for a thorough
review]{BartelmannSchneider:2001}.  WL can therefore calibrate the
relation between SZ observables and mass, and ultimately constrain
cosmology with SZ-selected clusters.

Thus there is a strong synergy between optical (including WL) and SZ
surveys. First, optical surveys enable a comparison between SZ and WL
signals and optical richness for the SZ-detected clusters. Second, a
multi-band optical imaging survey can reveal (photometric) redshifts for
SZ-detected clusters.  For these reasons, there are joint experiments
being planned: the Subaru Hyper Suprime-Cam (HSC) survey
\citep[][]{Miyazaki:2006}\footnote{\url{http://www.naoj.org/Projects/HSC/index.html};
also see \url{http://sumire.ipmu.jp/}} combined with the ACT survey, and
the Dark Energy Survey (DES;
\citealt{2005astro.ph.10346T})\footnote{\url{http://www.darkenergysurvey.org/}}
with the SPT survey.

With these upcoming SZ-WL surveys in mind, in this paper, we study WL
signal of a SZ-detected cluster, ACT-CL J0022.2$-$0036 (hereafter
ACTJ0022) at $z=0.81$, using multi-passband data with the current Subaru
prime-focus camera, Suprime-Cam \citep{Miyazaki:2002}. Subaru
Suprime-Cam is one of the best available ground-based instruments to
carry out accurate WL measurements, thanks to the excellent image
quality (median seeing FWHM is $0.6$--$0.7^{\prime\prime}$) and wide
field-of-view, $\sim 0.25$ deg$^2$
\citep{Miyazaki:2002b,Broadhurst:2005,Okabe:2010,Oguri:2012}. ACTJ0022
is one of the most luminous SZ clusters discovered in the 148-GHz ACT
map of 268 square degrees, which is a part of 500 square degrees in its
equatorial survey field taken in 2009 and 2010 \citep{Reese:2012,
Hasselfield:inprep} and overlaps with SDSS Stripe 82 field. Long-slit
follow-up spectroscopy at the Apache Point Observatory of the brightest
cluster galaxy (BCG) confirms the redshift of $z=0.81$
\citep{Menanteau:inprep}. To do the WL analysis, we analyze different
exposures simultaneously to model the shape of every galaxy, based on
the elliptical Gauss-Laguerre (EGL) shapelet method
\citep{Bernstein:2002,Nakajima:2006}. In the multi-exposure fitting, we
can keep the separate PSF of each exposure, and therefore keep the
highest-resolution PSF in the analysis, which is not the case for the
use of stacked images for the WL analysis. Furthermore, we use
photometric redshift (photo-$z$) information, derived from the stacked
images of Subaru $Br'i'z'Y$ data, in order to define a secure sample of
background (therefore lensed) galaxies. Thus, we combine shape
measurements and photo-$z$ information to study the mass of ACTJ0022,
which has not been fully explored in previous WL studies of
high-redshift clusters. Our study assesses the capability of
ground-based data for a WL study of high-redshift, SZ-detected
clusters. We also discuss the implications of our WL result for the
SZ-cluster mass scaling relations, and whether or not the estimated mass
of ACTJ0022 is consistent with the $\Lambda$CDM structure formation
model that is constrained by various cosmological data sets, using the
method in \cite{Mortonson:2011}.

This paper is organized as follows. In Section~\ref{sec:observations},
we describe the Subaru/Suprime-Cam follow-up observations. In
Section~\ref{sec:data_analysis}, we describe the data analysis including
data reduction, photo-$z$ estimation, and galaxy shape measurement. Then
we show the WL result for ACTJ0022, and discuss the systematic error
issues and the cosmological implication in Section~\ref{sec:results}.
Throughout this paper we use the AB magnitude system. Unless explicitly
stated, we adopt a flat $\Lambda$CDM cosmology with $\Omega_m = 0.27$
and $H_0 = 72{\rm km}\ {\rm s}^{-1}\ {\rm Mpc}^{-1}$.

\section{Observation}
\label{sec:observations}

\begin{figure*}
\includegraphics[width=13cm]{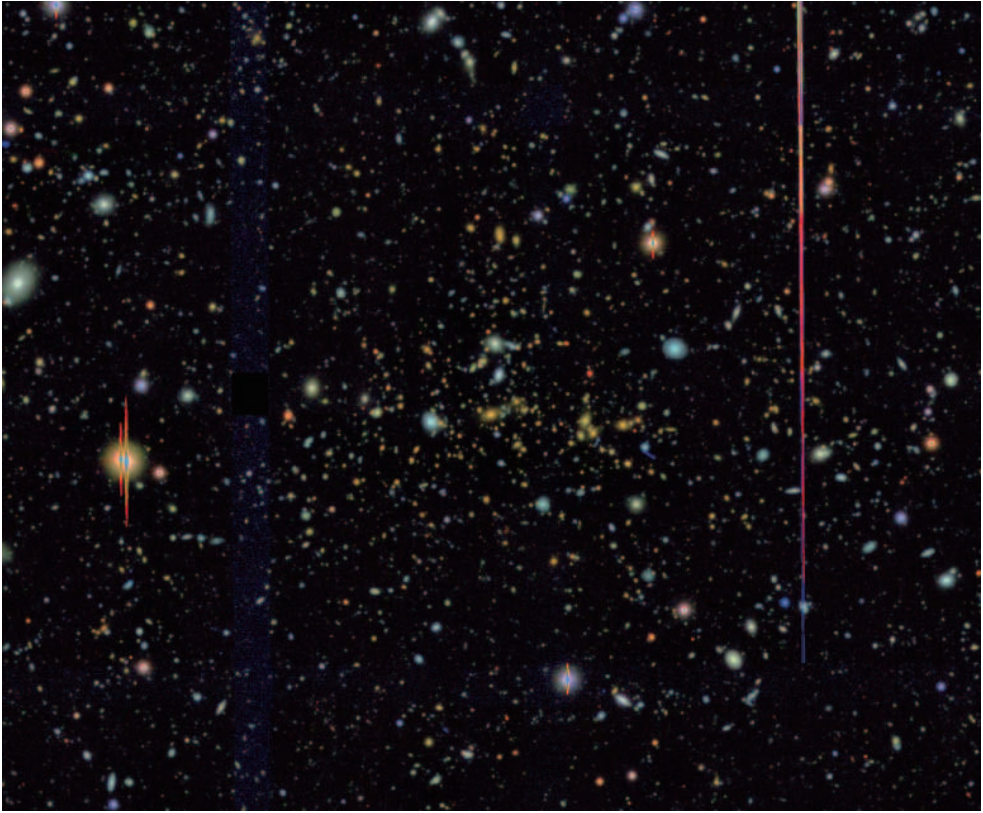}
\caption{ The Subaru/Suprime-Cam image of ACTJ0022, the region of about
$7\times 9$ square arcminutes around the cluster center (its BCG
position).  North is up and west is right. The color image is made by
combining the $r'i'z'$ images. Note that an angular scale of $1'$
corresponds to the transverse scale of 322~kpc$/h$ at $z=0.81$.}
\label{fig:cluster_RGB}
\end{figure*}

We observed the ACTJ0022 field on December 4th, 2010, using Suprime-Cam
\citep{Miyazaki:2002} with five broadband filters ($Br^\prime i^\prime
z^\prime Y$) on the Subaru Telescope \citep{Iye:2004}, as summarized in
Table~\ref{tab:observation}.  The RGB image of the cluster is shown in
Fig.~\ref{fig:cluster_RGB}. All passbands are used for photo-$z$,
whereas only the $i^\prime$-band image is used for shape
measurement. The choice of filters and depths was determined by using a
mock catalog of galaxies based on the methods of
\cite{Nishizawa:2010}. We constructed the mock catalog based on the
COSMOS photometric catalog \citep{Ilbert:2009}, and used the catalog to
estimate the required accuracy of photometric redshifts, available from
the multi-color data, in order to minimize contamination of foreground
and cluster-member galaxies (therefore unlensed galaxies) to the lensing
analysis.

\begin{table*}
\centering
\begin{tabular}{@{}cccccc@{}}
\hline \hline
filter & tot. exp. time [s] & \# of exp. & frame ID & typ. seeing [$^{\prime\prime}$] & lim. mag.\\
\hline
$B$ & 600 &3 & 1269250 - 1269279 & 0.66 & 25.9\\
$r^\prime$ & 600 & 3 & 1269680 - 1269709 & 1.06& 25.3\\
$i^\prime$ & 2400 & 10 & 1269320 - 1269419 & 0.74& 25.6\\
$z^\prime$ & 3240  & 12 & 1269430 - 1269549& 0.90& 24.8\\ 
$Y$ & 3240  & 12 & 1269560 - 1269679 & 0.78& 23.6\\
\hline
\end{tabular}
\caption{Summary of the Subaru/Suprime-Cam observations. Note that the
  limiting magnitude is for $3^{\prime\prime}$ aperture magnitude
  ($5\sigma$). $Y$-band is a 1 micron filter with the red edge
  defined by the deep-depletion CCD response.
\label{tab:observation}}
\end{table*}

\section{Data Analysis}
\label{sec:data_analysis}
\subsection{Analysis Overview}
\label{sec:analysis_overview}

\begin{figure*}
\includegraphics[width=10cm]{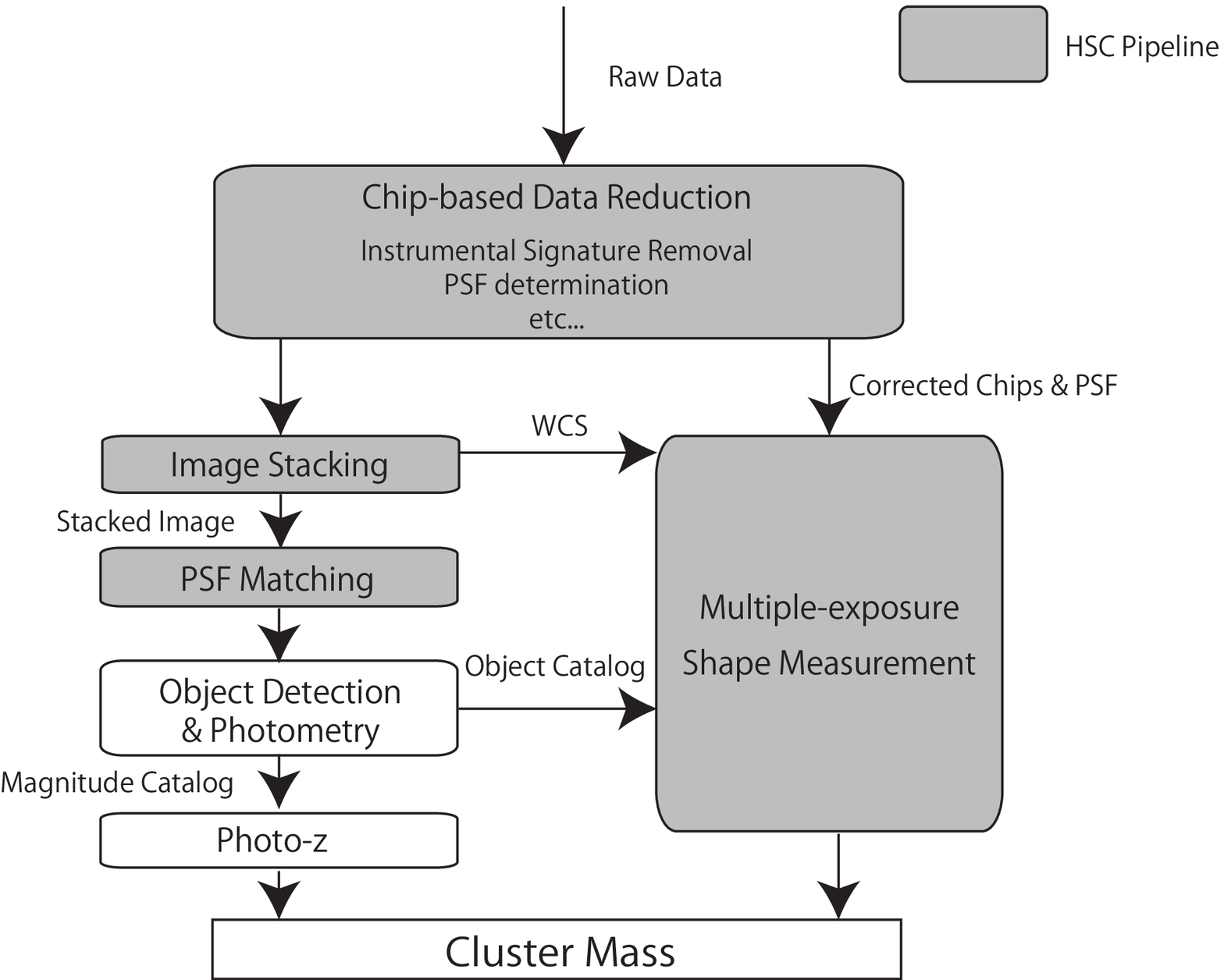}
\caption{Flow chart of our data analysis procedures.}
\label{fig:analysis_overview}
\end{figure*}

Figure~\ref{fig:analysis_overview} shows a flow chart of our data
analysis procedure.  In this analysis, we have used the HSC pipeline for
the tasks shown as shaded blocks.  The HSC pipeline is now actively
being developed for analysis of the HSC survey data, based on the data
reduction pipeline developed for LSST. Due to the large data volume (HSC
will provide $\sim2.3$GB per exposure), the pipeline aims to reduce the
data in an automated way from raw data to catalogs. Core parts of the
pipeline are written in C++ to enhance computing speed, then are wrapped
by a Python layer used to script together the core steps of the
analysis. We emphasize that our study is the first case where the HSC
pipeline is used for science. The version of the pipeline we use is
HSC.17.

The raw chip data first undergoes chip-based data reduction. At this
stage, instrumental signatures such as bias, overscan, and flat are
removed, and the PSF is determined (see Section~\ref{sec:data_reduction}
for details). The corrected chip data and PSF are passed to two
branches, for redshift determination and shape measurement of each
galaxy. To estimate photo-$z$, we stack all exposures for each chip,
match PSFs between different passbands, detect objects, carry out
photometry, and finally feed the measured magnitudes into the photo-$z$
software (see details in Section~\ref{sec:redshift_estimation}).

For the shape measurement, we employ the EGL method that aims to extract
shape information by representing the PSF and galaxy image with
orthogonal basis functions \citep{Bernstein:2002, Nakajima:2006}. We
analyze individual exposures simultaneously, which enables us to avoid
mixing PSFs taken in different epochs and interpolating pixel values.
Details of the shape measurement will be described in
Section~\ref{sec:shape_measurement}.  Finally, the photo-$z$ and shapes
are used for cluster mass estimation (Section~\ref{sec:results}).

\subsection{Chip-based Data Reduction}
\label{sec:data_reduction}

For each chip, the HSC pipeline produces three image planes with the
same dimensions (approximately 2k$\times$4k pixels). The first is an
image plane that contains the corrected image data. The second is a
variance plane that stores theoretical variance of each pixel; the noise
is first estimated from the raw image by assuming Poisson noise of
photon counts in each pixel, and then the noise is properly propagated
at each stage of the reduction. The third is a mask plane that has a
16-bit integer for each pixel. Different bits are used for different
masks to indicate saturation and other issues.

\subsubsection{Instrumental Signature Removal}

First, pixels having a value greater than a saturation threshold are
masked as {\tt SAT}. Different saturation thresholds are set for
each CCD according to its own characteristics.

A CCD has four outputs (or amps), each of which reads out
4177$\times$512 pixels. Thus the raw data of each CCD has four
stripes of image data, between which overscan regions are laid out.
Using the median of the overscan regions, the bias level is
subtracted. The overscan regions are then trimmed and the four stripes
are combined.

Now that we have signals only from photons, the variance plane is
created. Assuming Poisson statistics, the variance at pixel $(x,y)$ is
calculated as ${\rm Var}(x,y) = I(x,y)/g$, where $I(x,y)$ is the pixel
count in ADU and $g$ is the gain (the number of electrons per ADU). Note
that we use gains known for each amplifier independently.

Flat fielding and fringe correction are carried out using the dome flat
and sky frames, respectively. The CCD defects known beforehand are
masked as {\tt BAD}.  We also masked the pixels surrounding the
saturation masks by two additional pixels to avoid effects of electrons
leaking out from the saturated pixels.

We performed initial sky subtraction as follows (we refined the sky
subtraction at a later stage as we will describe below).  A chip image
is divided into patches, each of which contains 1024$\times$1024 pixels,
and the background in each patch is calculated using the 3-$\sigma$
clipping mean method. The background field is then obtained by
spline-interpolating the measured mean values at the center of each
patch. We then subtracted the background level in each pixel from the
image.

\subsubsection{Calibration}

We use bright sources to perform PSF measurement and astrometry.

First, we need to remove cosmic rays from the images. Assuming a
Gaussian PSF with FWHM 1.0$^{\prime\prime}$ as an initial guess, we
regard objects having sharper jump in the flux in one dimension and
smaller size than the PSF as a cosmic ray, and mask the associated
pixels as {\tt CR}.

We perform detection of bright objects as follows. By convolving the
image with the Gaussian PSF of $1.0^{\prime\prime}$ FWHM, we register a
set of connected pixels above the threshold value $n_{\rm th} \sigma$ as
a footprint of an object, where $\sigma$ is the sky noise and we
employed $n_{\rm th}=2$ in this analysis. Then we define bright objects
as a subset of objects with peak value above $n_{\rm th, ex}\times
n_{\rm th}\sigma$ in the original image, where we employed $n_{\rm th,
ex}=5$ (i.e. we adopted 10$\sigma$ for the peak value). At this step, we
again perform sky subtraction by using finer-size patches, each of which
has $128\times 128$ pixels, but masking the footprint of detected
objects. For the sky subtraction, we noticed that it is important to
mask outskirts of the detected objects; otherwise, the sky is
over-subtracted. This is the main reason we employed a rather
conservative value of $n_{\rm th}=2$ for the threshold value of object
detection.

Then we measured the PSF flux and second-order moments of each bright
object, using its image in the footprint.  The PSF flux $f_{\rm PSF}$ is
defined
by minimizing $\chi^2=\sum_\alpha^{N_{\rm pix}} [I_{\rm
data}(\bmf{x}_\alpha) - f_{\rm PSF}\hat{I}_{\rm
PSF}(\bmf{x}_\alpha)]^2/\sigma_\alpha^2$, where the index $\alpha$ runs
over the pixels of the footprint, $I_{\rm data}(\bmf{x}_\alpha)$ is the
image value at the $\alpha$-th pixel, $\sigma_\alpha$ is the noise at
the pixel, $\hat{I}_{\rm PSF}$ is the PSF function (the Gaussian
function of $1.0^{\prime\prime}$ FWHM up to this stage), and $f_{\rm
PSF}$ is a model parameter for the PSF flux. Note that the PSF profile
$\hat{I}_{\rm PSF}$ is normalized so as to satisfy $\sum_\alpha^{N_{\rm
pix}} \hat{I}_{\rm model}(\bmf{x}_\alpha) = 1$, and the center of the
PSF profile $\hat{I}_{\rm PSF}$ is set to the object center.  The
best-fit $f_{\rm PSF}$ is obtained by minimizing the $\chi^2$ above.
This is a linear algebra problem, so $f_{\rm PSF}$ can be obtained
without any ambiguity.  We also estimate the second-order moments of the
bright object, using adaptive moments defined as $M_{ij} = \int
W(\bmf{x})I(\bmf{x})x_i x_j d\bmf{x}$, where the integration runs over
all the pixels in the footprint and $W(\bmf{x})$ is a weight function.
We employed an elliptical Gaussian for $W(\bmf{x})$, whose shape is
matched to the object via an iterative
procedure\footnote{\url{http://www.sdss3.org/dr8/algorithms/classify.php#photo_adapt}}.

\subsubsection{PSF Determination}
\label{sec:PCA_PSF}

By using the PSF flux and the adaptive moments, we select star
candidates for PSF determination as follows.  We first remove objects
having the PSF flux below $f_{\rm lim}$ in order to eliminate faint,
small galaxies or low-$S/N$ stars.  In this analysis, we employ $f_{\rm
min}=60000$ counts corresponding to apparent magnitude brighter than
$\simeq 21.8$~mag.  We select star candidates from objects lying within
the 2$\sigma$ regions around the peak in the two-dimensional
distribution of $I_{11}$ and $I_{22}$, because stars should have small
moments and similar values. Since the variation of the second-order
moments is moderately large especially for the corner chip of the
Suprime-Cam focal plane, we decided to employ the 2$\sigma$ threshold,
rather than $1\sigma$, in order not to miss real stars in the
selection. Note that with this large $\sigma$ we can get compact
galaxies, which will be rejected by the following process.

Next, using the star candidates on each CCD chip, the PSF is
heuristically determined by principal component analysis \citep[PCA;
also known as Karhunen-Lo\`eve transform;][]{Jolliffe:1986}, with the
algorithm from the SDSS imaging pipeline \citep{Lupton:2001}. An image
of each star candidate can be represented by linear combination of
principal components (or eigenfunctions):
\begin{equation}
P(u,v) = \sum_{i=0}^{n_{\rm pc}-1}a_i K_i(u,v),
\end{equation}
where $P(u,v)$ is the observed image, $K_i(u,v)$ are the $i$-th
principal components, $n_{\rm pc}$ is a parameter to determine up to
which order principal component to include, and $u,v$ is the pixel
coordinates relative to the origin of principal components.  We include
the spatial variation of PSF assuming that the spatial variation of the
coefficients is modeled by the Chebyshev polynomials:
\begin{equation}
a_i \rightarrow a_i(x,y) \equiv \sum_{p=q=0}^{p+q \leq n_{\rm sv}} 
c_{pq}T_p(x)T_q(y),
\end{equation}
where $x,y$ is the pixel coordinates of a given CCD chip, $T_i(x)$ is
the $i$-th Chebyshev polynomial (employed to prevent the polynomial from
blowing up at the edge of chip), $c_{pq}$ is the expansion coefficients,
and $n_{\rm sv}$ is a parameter to determine which order of the
polynomials to include in this interpolation.  Note that the constraints
to determine the coefficients $c_{pq}$ are given at the positions of
stars, used for PSF determination, and the coordinates $(x,y)$ are
normalized to $[-1,1)$ across the chip for our convenience. In this
analysis, we set $n_{\rm pc}$ and $n_{\rm sv}$ to 6 and 4, respectively,
which are decided after an iterative, careful study of the PSF
determination (see Section~\ref{sec:shape_measurement_formalism} for
details).  The principal components $K(u,v)$ and the coefficients
$a_i(x,y)$ enable us to reconstruct the PSF at arbitrary positions,
which hereafter we refer to as the PCA PSF. Using the updated PSF
estimate, the PSF flux is re-measured for each bright object in order to
refine the star catalog (or remove the contaminating star-like
objects). After several iterations, we use the refined PSF estimates for
the update of cosmic ray masking and the following analysis.

\subsubsection{Astrometry}
\label{sec:astrometry}

The bright stars are matched to a reference catalog created from SDSS
DR8 \citep{Aihara:2011} by using {\tt
astrometry.net}\footnote{\url{http://astrometry.net/}}
\citep{Lang:2010}, which is the {\em astrometry engine} to create
astrometric meta-data for a given image. Based on the match list, we
determine the world coordinate system (WCS) in the TAN-SIP convention
\citep{Shupe:2005}. For this chip-based astrometry, we used quadratic
polynomials to obtain the transformation between the celestial
coordinates and pixel coordinates. The pixel scale of Subaru/Suprime-Cam
is about $0.2^{\prime\prime}$, which in fact slightly changes with
position due to the camera distortion. Note that we use the chip-based
WCS when co-adding different exposures to make the stacked images, and
then use the improved astrometry to renew the WCS for each chip.

\subsection{Redshift Estimation}
\label{sec:redshift_estimation}

 In this subsection, we describe the method for photometry, which will
then be needed for photo-$z$ estimation of galaxies (the left branch of
Fig.~\ref{fig:analysis_overview}).  Our method follows the prescription
proposed by \cite{Hildebrandt:2012}. A brief summary of our method of
determining galaxy photometry is: (1) stack (co-add) the corrected
images of each passband for detection of fainter objects, (2) match the
PSF across all the passband images, including PSF homogenization across
spatial positions, and (3) measure the aperture photometry of each
object, after the PSF matching, in order to robustly measure the colour
of objects for the same {\em physical} region. Several photometry
algorithms are now in development for the HSC pipeline. In this paper,
we decided to use \textsc{SExtractor} \citep{Bertin:1996} in order to
follow the method of \cite{Hildebrandt:2012}.  Below, we describe the
details of this procedure.

\subsubsection{Stacking and PSF Matching/Homogenization}
\label{sec:stacking_PSF_matcing} 

We stack different exposure images primarily by matching the positions
of stars, which are used for astrometry as described in
Section~\ref{sec:astrometry}, but also by matching slightly fainter
objects for a further improvement. The relative accuracy of our
astrometry is $\sim 0.03^{\prime\prime}$ (external + internal) and $\sim
0.01^{\prime\prime}$ (internal only). Here, ``external'' means accuracy
with respect to the external reference catalog and ``internal'' means
accuracy within the exposures we analyse. For the stacked image, WCS
based on the TAN-SIP convention is generated by using the matching list,
where we used the polynomials including terms up to $x^ny^m$, where
$n+m=10$ ($x,y$ are the pixel coordinates from the center of the stacked
image).  We use the celestial coordinates for the multiple-exposure
shape measurement as we will describe in
Section~\ref{sec:shape_measurement_formalism}.  When co-adding the
different images, we perform the scaling of each exposure based on the
measured PSF in each chip, such that the PSF fluxes (or the fluxes of
the same stars) in different exposures become identical.  The scaling
amplitude is typically within $1\pm0.02$.  Using the WCS and scaling
information, each exposure image is warped and the counts are scaled.
The warping requires resampling (or interpolation) of pixel values for
which we use the Lanczos3 algorithm to preserve independence of photon
noise in between different pixels\footnote{The sinc function is the
ideal interpolation, since it does not introduce any information whose
frequency is higher than the pixel sampling scale. However, because of its
infinite extent, we use a windowed approximation known as the Lanczos
filter.}.  Note that the resampling for all the $B r^\prime z^\prime Y$
images is matched to the $i^{\prime}$-band WCS, the details of which
will be described in Section~\ref{sec:photometry}.  After these
procedures, we stack all the exposures of a given passband.

To match PSFs of different passbands, we first find the largest PSF
among the stacked $Br^\prime i^\prime z^\prime Y$ images. We run the PSF
determination algorithm on each stacked image, and measure the adaptive
moments of the PCA PSFs at several spatial positions across the image.
The size of each PSF image is estimated from the adaptive moments as
\begin{equation}
\sigma = \left(M_{11}M_{22} - M_{12}^2\right)^{1/4} = |\mathrm{det}\,\, M|^{1/4}.
\end{equation}
The largest PSF we found is $\sim2.6$~pixels, around the edge of the
$r^\prime$-band stacked image.  For the PSF matching, we use the
algorithm developed by \cite{Alard:1998} and \cite{Alard:2000}
\citep[also see][for the recent implementation]{Huff:11}.  This method
enables us to match the PSFs to an arbitrary, analytical PSF shape, the
so-called target PSF, by convolving the observed image with the
differential PSF kernel.  The target PSF we use in this analysis is the
Gaussian function, a convenient approximation to PSF, with
$\sigma=2.6$~pixels matching the largest PSF above. Furthermore, we
implement homogenization of the matched PSF across the spatial positions
in the image; i.e., we use a spatially-varying kernel in order to have
the same PSF across all the positions in the matched image.

Table~\ref{tab:PSFmatch} shows the size and ellipticity of PSFs before
and after the PSF matching, where the error shows the standard deviation
of the quantities across the field and the ellipticity is estimated from
the adaptive moments as
\begin{equation}
\left(e_1,e_2\right) = \left(\frac{M_{11} - M_{22}}{M_{11} + M_{22}},\frac{2M_{12}}{M_{11} + M_{22}} \right).
\end{equation}
The PSF size in each band is matched to 2.6 pixels within about 1.5 per
cent, and the ellipticity of the matched PSF is consistent with zero.

\begin{table}
\begin{center}
\begin{tabular}{|ccccc|}
\hline \hline
\multicolumn{2}{|c|}{} & $\sigma$ [pixel] & $e_1$ & $e_2$ \\ \hline
$B$ & original & 1.39 $\pm$ 0.04 & 0.055 $\pm$ 0.023 & -0.002 $\pm$ 0.013 \\ \cline{2-5}
& match & 2.61 $\pm$ 0.02 & -0.003 $\pm$ 0.005 & -0.001 $\pm$ 0.004 \\ \hline
$r^\prime$ & original & 2.28 $\pm$ 0.11 & -0.032 $\pm$ 0.018 & -0.009 $\pm$ 0.019 \\ \cline{2-5}
& match & 2.57 $\pm$ 0.06 & -0.001 $\pm$ 0.011 & -0.002 $\pm$ 0.012 \\ \hline
$i^\prime$ & original & 1.55 $\pm$ 0.05 & -0.019 $\pm$ 0.025 & -0.006 $\pm$ 0.035 \\ \cline{2-5}
& match & 2.61 $\pm$ 0.03 & 0.001 $\pm$ 0.008 & -0.002 $\pm$ 0.012 \\ \hline
$z^\prime$ & original & 1.91 $\pm$ 0.06 & -0.015 $\pm$ 0.020 & -0.022 $\pm$ 0.026 \\ \cline{2-5}
& match & 2.60 $\pm$ 0.04 & 0.001 $\pm$ 0.009 & -0.003 $\pm$ 0.013 \\ \hline
$Y$ & original & 1.65 $\pm$ 0.08 & 0.000 $\pm$ 0.025 & -0.023 $\pm$ 0.035 \\ \cline{2-5}
& match & 2.60 $\pm$ 0.04 & 0.001 $\pm$ 0.008 & -0.002 $\pm$ 0.012 \\ \hline
\end{tabular}
\end{center}
\caption{ The PSF size and average ellipticity for each passband stacked
images. The row labelled as ``original'' or ``match'' shows the results
for the stacked images with or without the PSF match/homogenization (see
Section~\ref{sec:stacking_PSF_matcing} for details).  }
\label{tab:PSFmatch}
\end{table}

\subsubsection{Photometry}
\label{sec:photometry}

 We use \textsc{SExtractor} to
perform object detection 
as well as photometry for the PSF-matched, stacked images. As
we stressed, we want to measure the flux of each object for the {\em
same} region (and with the {\em same} weight).  First, we use the
stacked $i'$-band image, before the PSF matching, for object detection
as well as for defining the photometry region, because the images before
the PSF matching are higher resolution 
and are less contaminated by the blending of neighboring
objects. For the photometry region, in this analysis, we use the
isophotal region around each object; we defined the group of connected
pixels around each object, which have counts above 5 times the sky
noise.  We can obtain this group
of pixels, called the {\em segmentation}
region, using \textsc{SExtractor}; it is conceptually equivalent to the footprint in the HSC
pipeline. Then we define the {\em same} photometry regions in the
stacked $Br'z' Y$ images by matching the segmentation region in the
$i'$-band image to the other passband image via the WCS, as described in
Section~\ref{sec:stacking_PSF_matcing}. After these procedures, we
finally make the aperture magnitude {\tt MAG\_ISO}, within the same
segmentation region, for each object in each of the PSF-matched, stacked
$Br'i'z' Y$ images, using the dual mode
of \textsc{SExtractor}, as suggested in \cite{Hildebrandt:2012}. 

To determine the magnitude zero point, we identify the SDSS stars in the
ACTJ0022 field and measure the star flux in a $4.8^{\prime\prime}$
aperture on the PSF-matched images.  We employ such a larger aperture in
order to cover all the flux from stars smeared by the PSF matching.
Although the SDSS DR8 photometry is calibrated at high precision
\citep{Aihara:2011}, we cannot directly compare the stellar fluxes
inferred from the SDSS catalog with the measured fluxes of the
Suprime-Cam data, because the $r'i'z'$ filter responses are not exactly
the same, and the $B$- and $Y$-passbands do not exist in the SDSS
photometric system.  Thus we need to infer the Suprime-Cam filter
magnitudes for each star from the SDSS magnitudes using the following
method.  First, we fit the multi-band fluxes in $ugriz$ for each stellar
object in the SDSS catalog to a stellar atmosphere model from
\cite{Castelli:2004}. The model includes 3808 stellar spectra that are
given as a function of various combinations of metallicities, effective
temperatures, and surface gravity strengths. By convolving the best-fit
spectrum with the response functions of the Suprime-Cam filters, we can
estimate the Suprime-Cam filter magnitudes for each SDSS star.  Note
that the Suprime-Cam $B$-band magnitude is effectively interpolated
between the SDSS passbands, whereas the $Y$-band magnitude is
extrapolated from the SDSS magnitudes.  Since the SDSS magnitudes are
already calibrated for atmospheric extinction at a reference airmass of
1.3, we do not have to correct for the airmass difference between
exposures. Using the above method, we determine the magnitude zero point
of each band. The errors of the zero point are estimated from the
scatters between the SDSS- and Suprime-Cam magnitudes as $B$: 0.048,
$r^\prime$: 0.090, $i^\prime$: 0.043, $z^\prime$: 0.080, and $Y$: 0.086
magnitudes.

We correct for Galactic dust extinction following the approach in
\cite{Schlegel:1998} and the dust extinction map provided by the
NASA/IPAC Infrared Science
Archive\footnote{\url{http://irsa.ipac.caltech.edu/applications/DUST/}}. The
estimated extinctions ($B$: 0.098, $r^\prime$: 0.066, $i^\prime$: 0.050,
$z^\prime$: 0.036, and $Y$: 0.031) are used to correct our photometry.

\subsubsection{Photometric Redshift}
\label{sec:photometric_redshift} 

For the photo-$z$ estimate, we use the publicly available code, {\em Le
Phare}\footnote{\url{http://www.cfht.hawaii.edu/~arnouts/LEPHARE/lephare.html}}
\citep{Arnouts:1999,Ilbert:2006}, which is based on template-fitting of
the galaxy spectral energy distribution (SED). The template set of SEDs
that we use is based on the CWW \citep{Coleman:1980} and starburst
templates \citep{Kinney:1996}.  The CWW templates were refined in order
to better match the actual data from the CFHTLS as well as the VVDS
spectroscopic data \citep{Ilbert:2006}.  In addition, {\em Le Phare} has
a functionality to re-calibrate magnitude zero points so that the
difference between the observed and model SEDs are adjusted using a
training set of spectroscopic galaxies.  In this analysis, we use
spectroscopic galaxy catalogs from the SDSS DR8 \citep{Aihara:2011} and
the Baryon Oscillation Spectroscopic Survey (BOSS;
\citealt{Eisenstein:2011,Bolton:2012,Dawson:2012,Smee:2012}).  For the
ACTJ0022 field, we have 205 spectroscopic redshifts from the catalogs to
use for the calibration.  The offsets of the magnitude zero points
obtained from this procedure are $B$: 0.072, $r^\prime$: 0.057,
$i^\prime$: -0.023, $z^\prime$: -0.053, and $Y$: 0.016, which are
comparable to the zero point errors shown in
Section~\ref{sec:photometry}.

\begin{figure*}
\includegraphics[width=8cm]{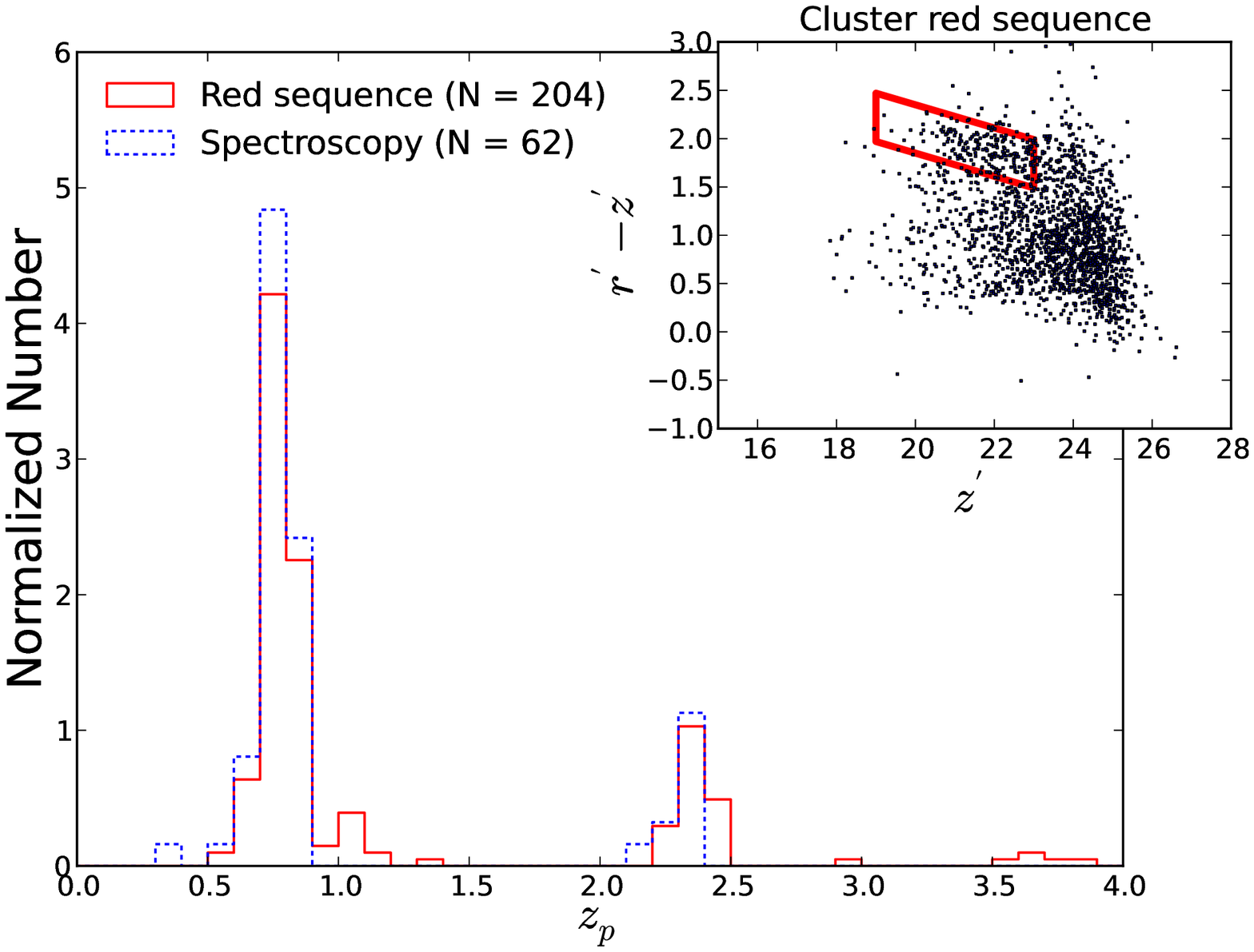}
\includegraphics[width=8cm]{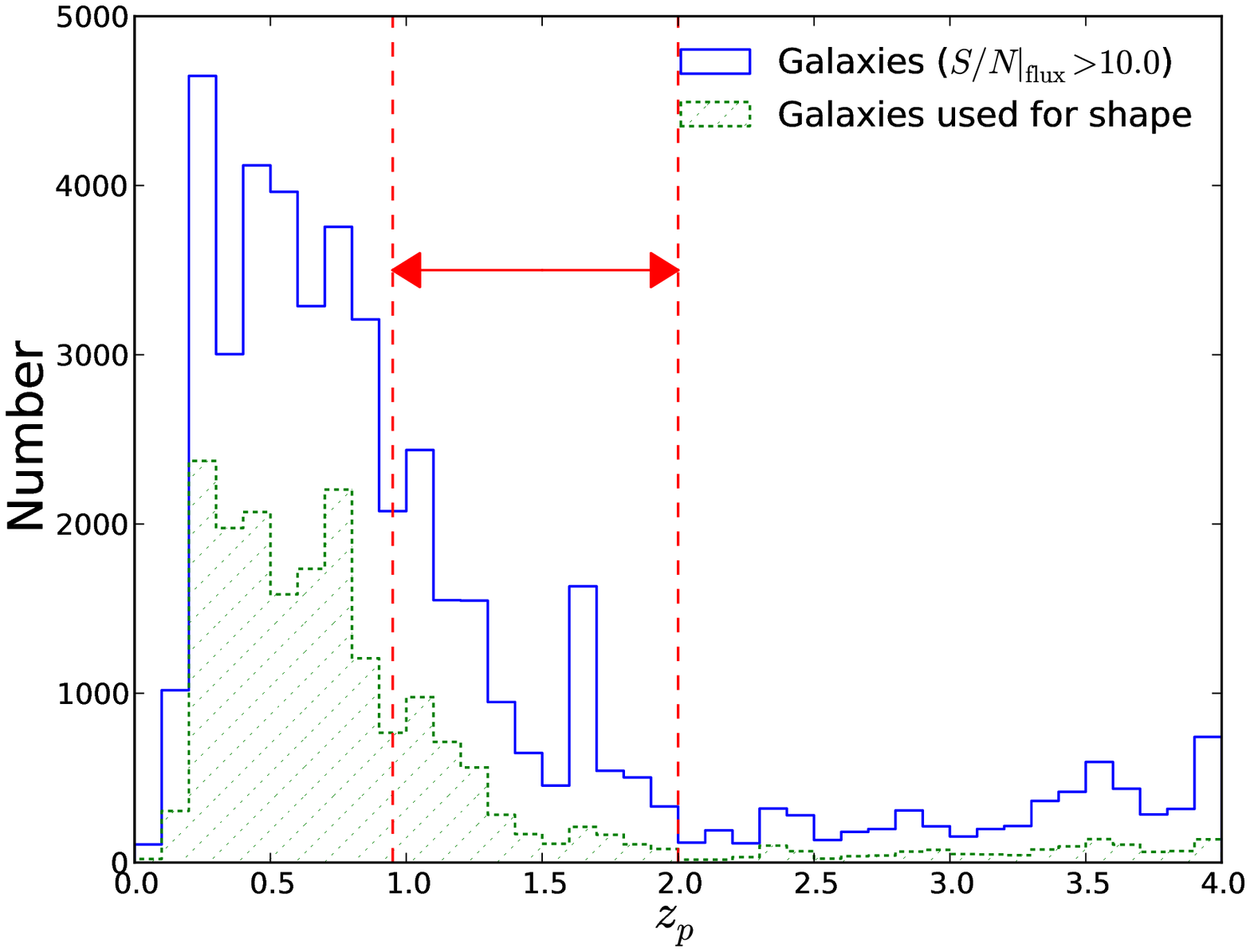} 
\caption{{\em Left panel:} The solid-line histogram shows the
distribution of the selected red-sequence galaxies in the ACTJ0022
field, as a function of their photometric redshift estimates $z_p$
($x$-axis). The red-sequence galaxies are selected by the solid-line box
in the color-magnitude diagram, as shown in the inset plot. For
comparison, the dotted-line histogram is the distribution of the
confirmed cluster members, again as a function of our photo-$z$
estimates of the galaxies, where the spectroscopic redshifts of cluster
members are taken with Gemni/GMOS and confirmed to be at the same
redshift of the cluster within 5000~km/s. Note that the amplitudes of
the histograms are normalized so that $\sum_i^{\rm bins} N_{z_p,i}\Delta
z = 1$, where $\Delta z = 0.1$. These photo-$z$s are consistent with the
cluster redshift of $z=0.81$, although there are some catastrophic
failures at $z>2.0$.  {\em Right panel:} The solid-line histogram shows
the photo-$z$ distributions of all the imaging galaxies that have $S/N>
10$ for the $3^{\prime\prime}$ aperture flux in their stacked $i^\prime$
images. The dotted-line and shaded histogram is the photo-$z$
distribution for the galaxies used for the weak lensing analysis, where
the size and flux cut are imposed on those galaxies to have a reliable
shape measurement. In our weak lensing analysis, we further impose the
photo-$z$ cut $0.95<z_p<2.0$, which is denoted by dashed vertical lines
and a solid arrow, to minimize contamination from the photo-$z$ outliers
indicated in the left panel.}  \label{fig:n_zp}
\end{figure*}

As one validation of our photo-$z$s, the left panel of
Fig.~\ref{fig:n_zp} shows the photo-$z$ distribution for galaxies
selected around the red sequence in the color-magnitude diagram, which
are therefore likely to be cluster members.  To be more precise, we
employ the red sequence given by the ranges in $19 < z^\prime < 23$ and
$-0.12z^\prime + 4.25 < r^\prime - z^\prime < -0.12z^\prime + 4.75$.  In
addition, we focus on the red galaxies located in a 2000$\times$2000
pixels region, or 6.7$\times$6.7 arcmin$^2$, around the BCG (i.e. a
proxy for cluster center), because a typical virial radius for a massive
cluster is about 2~Mpc, which corresponds to about 1300 pixels at
redshift $z=0.8$ for a $\Lambda$CDM model.  After imposing these
selection criteria, we find 238 red-sequence galaxies.

The figure compares our photo-$z$ estimates for the red galaxies with
spectroscopically-selected member galaxies, which were taken using
Gemini-south/GMOS (Program: GS-2011B-C-1, PI: F.~Menanteau) as a part of
the spectroscopic follow-up of ACT-SZ selected clusters
\citep{sifon:2012}.  Note that the Gemini spectroscopic galaxies shown
here are all the member galaxies, within 5000~km~s$^{-1}$ with respect
to the cluster.  The distribution of photo-$z$ for these confirmed
cluster member galaxies shows that there is a significant overlap of the
photo-$z$s with the spectroscopic redshifts around the cluster redshift
$z=0.81$.  However, the figure also shows that there are some
catastrophic failures of the photo-$z$s around $z_p\simeq 2.3$ and 3.8.
If we ignore these catastrophic photo-$z$ failures, the mean redshift of
the photometric red-sequence galaxies is $0.79 \pm 0.09$, which is in
good agreement with the cluster redshift within the error bars.  In the
following analysis, we conservatively use galaxies with photo-$z$s
$0.95<z_p<2.0$ as the catalog of background galaxies.  Thus we do not
use the galaxies with $z_p>2$, because the figure implies that the
catalog can be contaminated by unlensed member or foreground galaxies,
which cause a dilution of the estimated lensing signals
\cite[e.g.,][]{Broadhurst:2005}.

The right panel of Fig.~\ref{fig:n_zp} shows the redshift distributions
of photometric galaxies. The solid-line histogram is the photometric redshift
distribution for galaxies that have $S/N> 10$ for the
$3^{\prime\prime}$ aperture flux. The shaded histogram shows the redshift
distribution for galaxies that are useful for weak lensing analysis;
the galaxies have sufficiently large size and flux $S/N$ for the shape
measurement, as we will discuss in more detail in
Section~\ref{sec:shape_measurement_selection_criteria}.
Again, for the following lensing analysis, we use galaxies with
$0.95<z_p<2.0$ to minimize the contamination by photo-$z$
outliers. 

\subsection{Shape Measurement}
\label{sec:shape_measurement} 

For the shape measurement (the right branch of
Fig.~\ref{fig:analysis_overview}), we employ the EGL method which uses
elliptical Gauss-Laguerre (GL) basis functions to model galaxy images.
We also expand the method to simultaneous multiple-exposure measurement
to avoid mixing different PSFs in different exposures as well as pixel
resampling, which are systematic issues when using stacked images for
the shape measurements.

\subsubsection{Star-galaxy separation}
\label{sec:shape_measurement_selection_criteria}

As described in Section~\ref{sec:photometry}, we use the $i'$-band
stacked image for object detection as well as for star-galaxy
selection. Again note that we use the $i'$-band images for the shape
measurement.  We use the size-magnitude diagram to select stars from the
locus of objects with nearly constant FWHM and $19.5 < i^\prime < 21.5$,
yielding about $ 650$ stars in total, with mean size ${\rm
FWHM}=0.69^{\prime\prime} \pm 0.03^{\prime\prime}$.

To select galaxies, we use objects that have FWHMs more than $2\sigma$
above the stellar FWHM, where $\sigma$ is the stellar size rms. At this
stage, the number density is 52.7~arcmin$^{-2}$. We then applied the
magnitude cut $19 < i^\prime < 25.6$, where the faint end of the
magnitude range is determined so that the total signal-to-noise ratio
($S/N$) for the $3^{\prime\prime}$ aperture flux should be greater than
5. The number density is reduced to 48.6~arcmin$^{-2}$.  Together with
the photo-$z$ cut (see Section~\ref{sec:photometric_redshift}), the
resulting number density of source galaxies is about
10.6~arcmin$^{-2}$. Furthermore, after imposing size and $S/N$ cuts for
reliable shape measurements of galaxies, the final number density
becomes 3.2~arcmin$^{-2}$ (see
Section~\ref{sec:shape_measurement_galaxy_shape} for details).

\subsubsection{PSF fitting}
\label{sec:shape_measurement_formalism} 

First, we need to model the PCA-reconstructed PSF at the position of
each galaxy in each exposure, based on the GL eigenfunction
decomposition.  Note that, as we described in
Section~\ref{sec:photometry}, every galaxy is detected in the stacked
image, and the galaxy position was first defined in the pixel
coordinates of the stacked image. The coordinate transformation between
the pixel coordinates of the stacked image and a given exposure image is
given via the WCS, which is provided by the HSC pipeline.  The
coordinate transformation differs for the different exposures.  Hence we
perform the PSF modeling in the celestial coordinates ; the model for
the PCA PSF at the galaxy position and for the $\eta$-th exposure image
is given as
\begin{equation}
I^{\rm \ast(\eta)}\!(\bmf{\theta}^{(\eta)}) = \sum_{p,q} b_{pq}^{\ast(\eta)}
 \psi^{\sigma_\ast^{(\eta)}}_{pq}\!\!\!\left(\bmf{{\cal
			 W}}^{(\eta)}\!\!\left(\bmf{\theta}^{(\eta)}-\bmf{\theta}^{(\eta)}_0\right)\right),
\label{eq:GL_PSF}
\end{equation}
where $\psi^{\sigma_\ast^{(\eta)}}_{pq}\!\!(\bmf{\theta})$ is the
two-dimensional (circular) GL function with the order ($p,q$);
$\sigma_\ast$ is a parameter to determine the width of the GL functions;
$\bmf{b}^{\ast(\eta)}$ is the expansion coefficients; the operation
$\bmf{W}^{(\eta)}(\bmf{\theta}-\bmf{\theta}_0)$ transforms the pixel
coordinate in the $\eta$-th exposure to the celestial coordinates;
$\bmf{\theta}_0$ is the centroid of the PSF.  Thus, by modeling the PSF
in the celestial coordinates, we properly correct for the astrometric
distortion effect, which is treated as a coordinate transformation, not
a convolution effect, e.g. in the case for the atmospheric smearing
effect (the major part of PSF).

The fitting parameters of Eq.~(\ref{eq:GL_PSF}) are $(b^\ast_{pq},
\sigma_\ast, \bmf{\theta}_0)$. We employ the $\chi^2$ fitting via
$\chi^2=\sum_{\alpha}[I_{\rm
data}^\ast(\bmf{\theta}_\alpha)-I^{\ast}(\bmf{\theta}_\alpha)]^2/
\sigma^2_\alpha$, to determine the model parameters. The $\chi^2$
minimization with respect to the parameters $b_{pq}^\ast$ can be reduced
to a linear algebra problem, so $b_{pq}^\ast$ can be uniquely determined
for given $\sigma_\ast$ and $\bmf{\theta}_0$, thanks to the
orthogonality of the eigenfunction.  Hence we need to find the best-fit
$\sigma_{\ast}$ and $\bmf{\theta}_0$ by minimizing the $\chi^2$-value,
at the galaxy position in each exposure.

\begin{figure*}
\includegraphics[width=5.5cm]{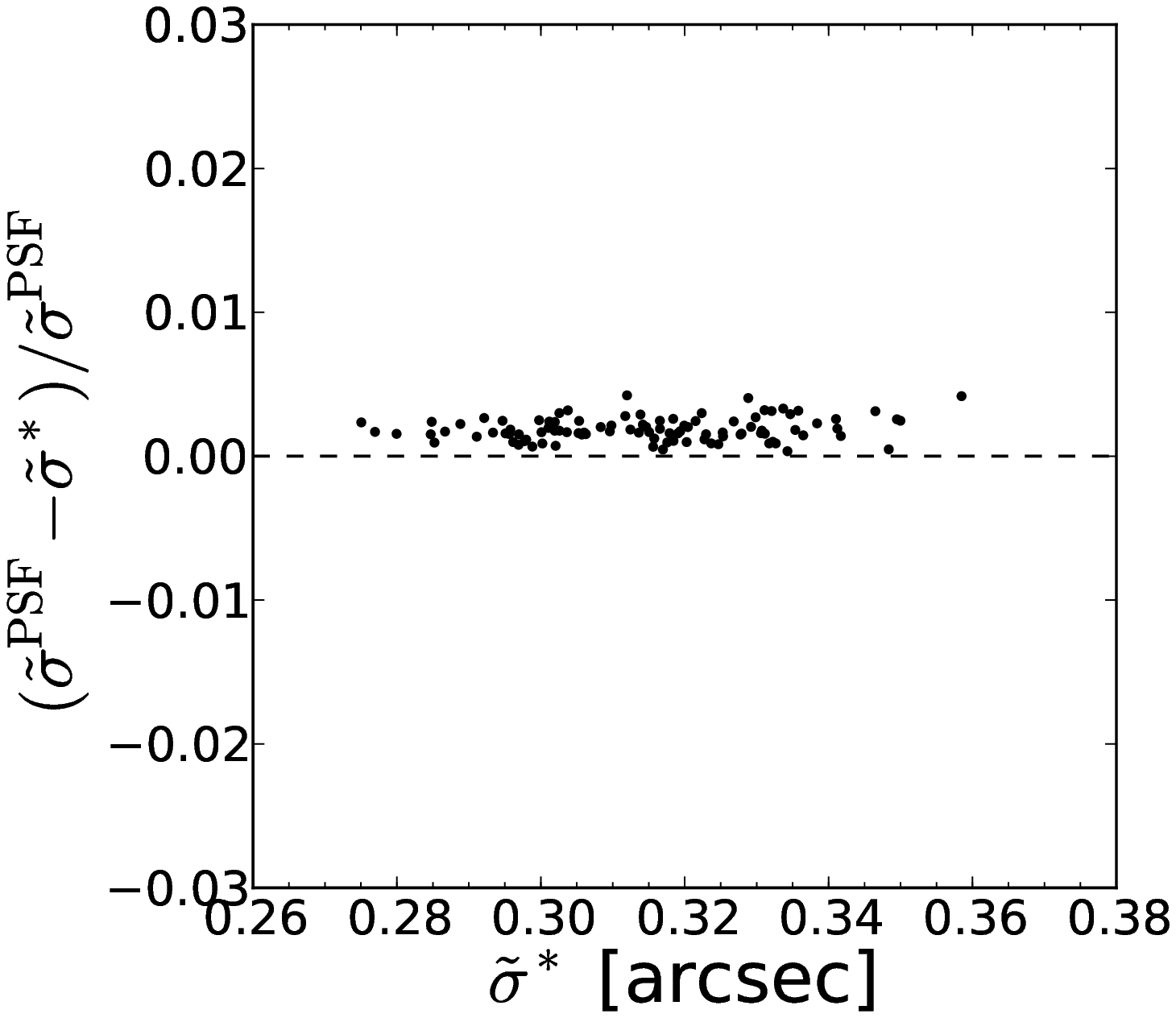}
\includegraphics[width=5.5cm]{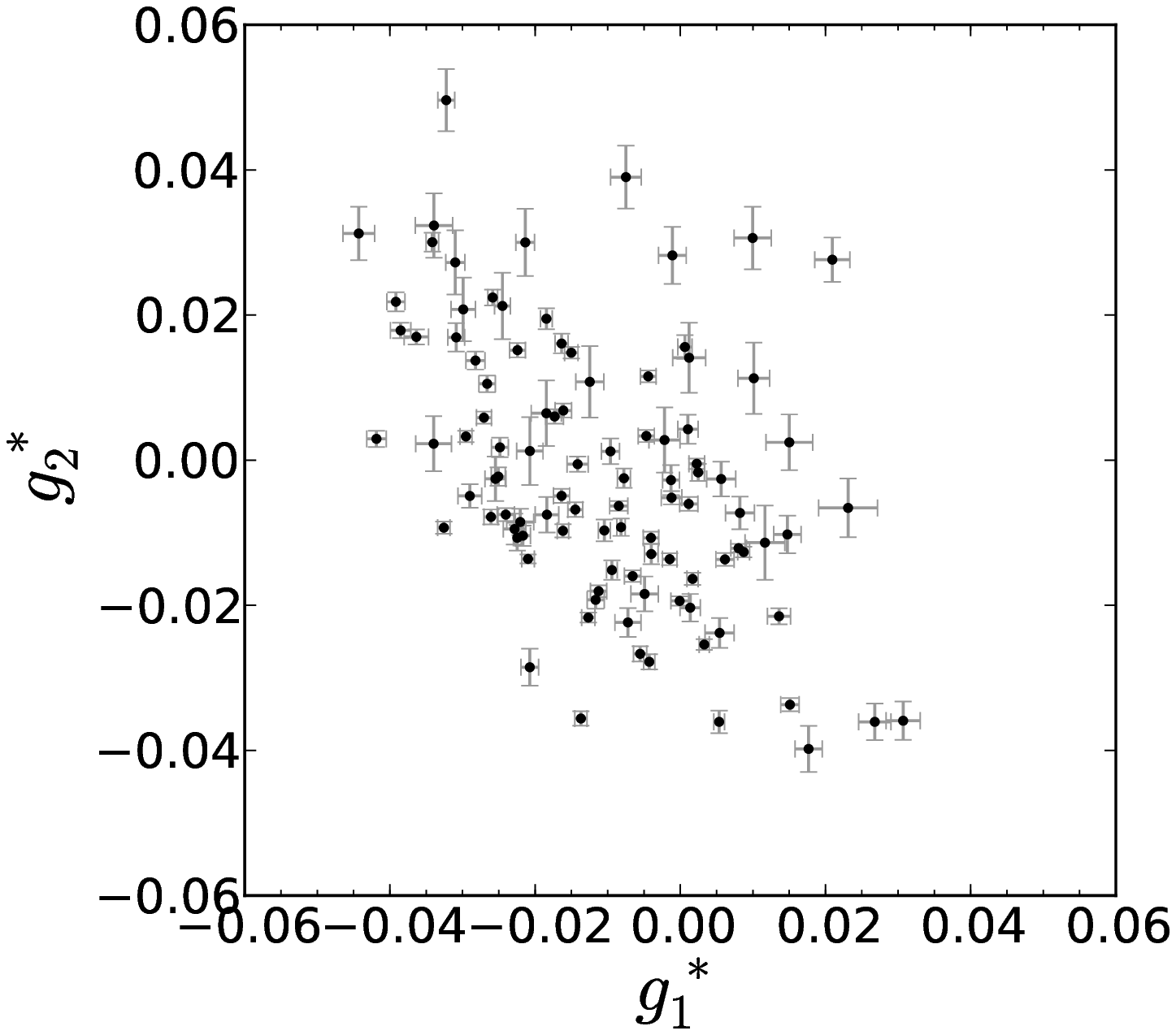}
\includegraphics[width=5.5cm]{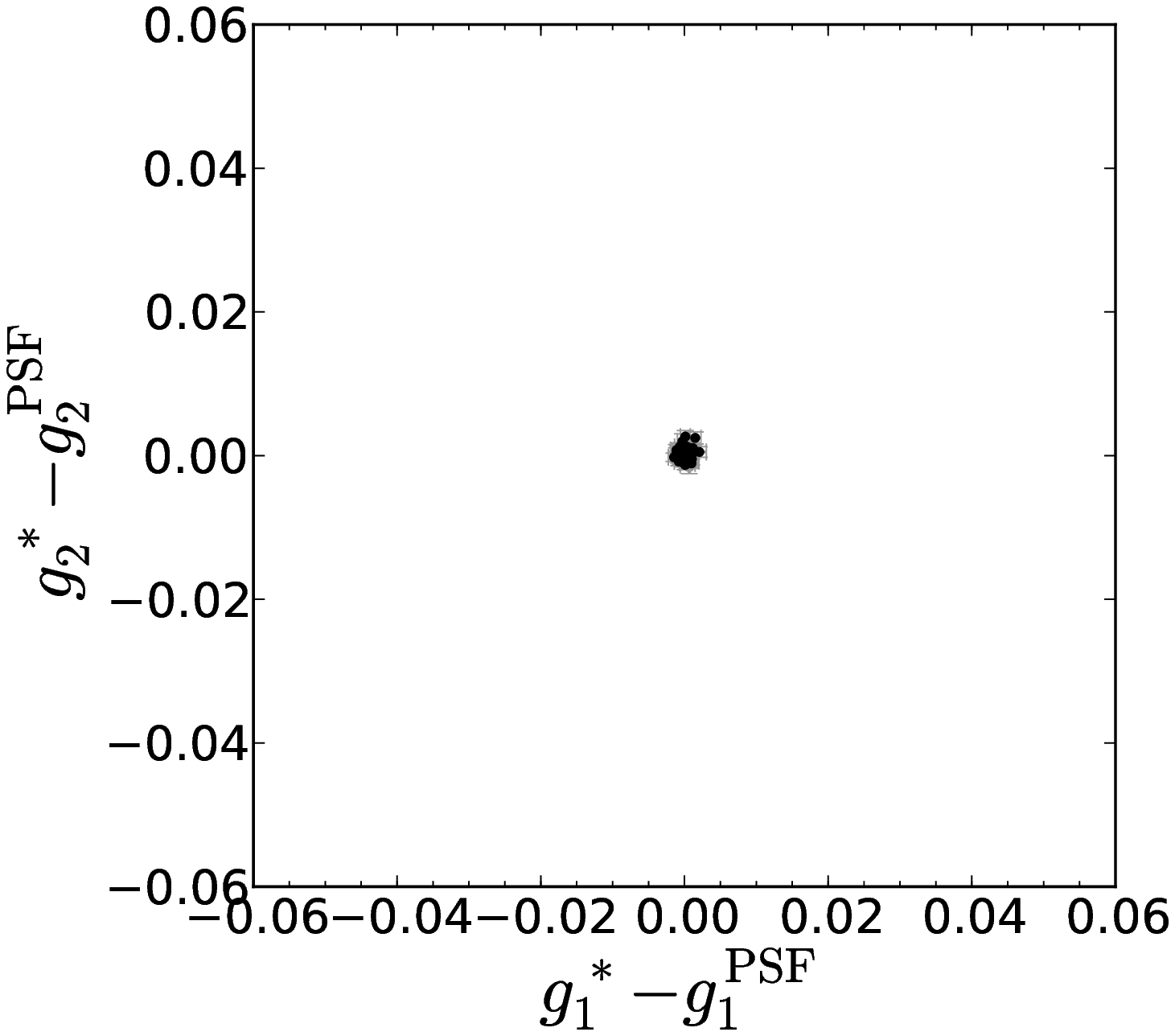}
\caption{ {\em Left panel: } The fractional differences between the
sizes of the PCA-reconstructed PSFs and star images as a function of the
star size (see Eq.~\ref{eq:bcoeff_size_ellipticity} for the size
definition).  Each dot denotes the mean values measured from each CCD
chip (we have 100 results in total, 10 CCD chips times 10 exposures).
The dashed line denotes the relation $\tilde{\sigma}^{\rm PSF} =
\tilde{\sigma}^\ast$. {\em Middle panel: } The measured ellipticities of
stars in each CCD.  Each dot with error bars denotes the mean value of
the ellipticities of stars lying in a given chip, and the error bars are
the standard deviation.  {\em Right panel: } Similar to the middle
panel, but for the residual ellipticities between the stars (in the
middle panel) and the PCA-reconstructed PSFs. Here, in each chip, we
measured ellipticities of the PCA-reconstructed PSF at each star
position, subtracted the observed star ellipticity, and computed the
mean and standard deviation (see Section~\ref{sec:PCA_PSF} for the
PCA-PSF determination).  Note that we measured the size and
ellipticities using the Gauss-Laguerre shapelet method (see
Section~\ref{sec:shape_measurement_selection_criteria}).  }
\label{fig:ellipticity_calibrate}
\end{figure*}

As an estimate of the accuracy of our PSF measurement, we compare the
size and ellipticities of each star image with 
those of the
PCA-reconstructed PSF image at the star position. 
Using the best-fit $\bmf{b}^\ast$
coefficients, the size and conformal shear of objects can be estimated  
\citep{Bernstein:2002} as
\begin{eqnarray}
\tilde{\sigma}^\ast &=& \sigma^\ast \exp{\left(\frac{b_{11}^\ast}{b_{00}^\ast
 - b_{22}^\ast}\right)},\nonumber \\
\bmf{\eta} &=& \frac{2\sqrt{2}b_{02}^\ast}{b_{00}^\ast-b_{22}^\ast}.
\label{eq:bcoeff_size_ellipticity}
\end{eqnarray}
Then we convert $\bmf{\eta}$ to the reduced shear as $g = {\rm
tanh}\left( \eta/2 \right)$.  Fig.~\ref{fig:ellipticity_calibrate} shows
the results for this comparison. Note that we performed the same fitting
described in the earlier part of this section for each star image to
obtain the best-fit $\bmf{b}^\ast$ coefficients.  The fractional size
difference between the PCA-PSF and star sizes agrees to within 0.2 per
cent.  The typical residual of ellipticities on each chip is
$\bmf{g}^\ast - \bmf{g}^{\rm PSF} = (1.4\pm 6.5, 0.6\pm 6.4)\times
10^{-4} $ (the mean and RMS in the chip averaged over the different
exposures), and is consistent with zero.  These residuals would
contaminate galaxy shapes as an additive bias.  We will discuss the
impact on the cluster mass estimation from the measured weak lensing
signal in Section~\ref{sec:results_systematics_measurement}.

\subsubsection{Galaxy shape measurement}
\label{sec:shape_measurement_galaxy_shape} 

For simultaneous multi-exposure fitting of a given galaxy shape, we use
the same model parameters for different images in different exposures.
Note that the internal astrometric errors are typically
$\sim0.01^{\prime\prime}$, as describe in
Section~\ref{sec:stacking_PSF_matcing}.  Hence we believe that the
coordinate transformations between different exposures are known
accurately enough, and the astrometric errors do not induce a
significant systematic error in the lensing shear estimate (less than 1
per cent; see \citealt{Miyatake:inprep}).

In our fitting procedure, we first estimate a size for the PSF-convolved
image of a given galaxy, by combining the different exposures based on
the GL eigenfunction decomposition:
\begin{flalign}
&\chi^2 = \sum_{\eta}^{N_\mathrm{exp}} \sum_\alpha^{N_{\rm
 pix}^{(\eta)}} &\nonumber \\
&\frac{\left[f^{(\eta)}_s
I^{\rm (\eta)}\hspace{-0.2em}(\bmf{\theta}_\alpha^{(\eta)}) -
		       \sum_{p,q} b_{pq}^{(\eta)}
 \psi^{\sigma_o^{\rm ini} E}_{pq}\!\!\left(\bmf{{\cal
		      W}}^{(\eta)}\!\!\left(\bmf{\theta}_\alpha^{(\eta)}-\bmf{\theta}^{(\eta)}_0\right)\right)\right]^2}{\left(f^{(\eta)}_s\sigma_\alpha^{(\eta)}\right)^2}, &
\label{eq:chisq_native_fit}
\end{flalign}
where $\alpha$ runs over pixels in the segmentation region around the
galaxy (see Section~\ref{sec:photometry}); $\sigma_\alpha^{(\eta)}$ is
the sky noise at the position $\bmf{\theta}_\alpha$ of the $\eta$-th
exposure; $f_s^{(\eta)}$ is the scaling factor of the exposure estimated
by the HSC pipeline (Section~\ref{sec:stacking_PSF_matcing}); and
$\psi_i^{\sigma_{\rm o}^{\rm ini}E}$ are {\em elliptical} GL functions
that have width $\sigma_{\rm o}^{\rm ini}$, for which we use
$\sigma_{\rm o}^{\rm ini}=1.49$~pixels as the initial guess.  Following
the method in \cite{Nakajima:2006}, a galaxy image is modelled in a
sheared coordinate system rather than in the sky plane, because the
lensing shear distortion is equivalent to an elliptical coordinate
transformation. More precisely, the elliptical GL functions are defined
as
\begin{eqnarray}
\psi_{pq}^{\sigma E}(\bmf{\theta}) &\equiv& 
\psi^{\sigma}_{pq}\!(\bmf{E}^{-1}\bmf{\theta}), \\
\bmf{E}^{-1} &\equiv& \frac{e^{-\mu}}{\sqrt{1-g^2}}\left(
\begin{array}{cc}
1-g_1 & -g_2 \\
-g_2 & 1+g_1 \\
\end{array}
\right).
\label{eq:GL_coord_transformation}
\end{eqnarray}
Here \bmf{E} represents a coordinate transformation from the sky plane
that includes a two-dimensional translation, a shear $\bmf{g}$, and a
dilution $\mu$.  There are 5 fitting parameters, $(\mu, g_1, g_2, x_c,
y_c)$, where $(x_c, y_c)$ is the centroid position of the
galaxy. Following \cite{Nakajima:2006}, we minimize $\chi^2$ so that the
obtained coefficients $b_{pq}$ satisfy the so-called ``null test'' given
by $b_{10} = b_{01} = b_{11} = b_{20} = b_{02} = 0$.  This
$\chi^2$-minimization gives an estimate of the size of the {\em
observed} galaxy as $e^{\mu} \sigma_{\rm o}^{\rm ini}$,which includes
the PSF smearing effect. We define $\sigma_{\rm gal} = \left(e^{\mu}
\sigma_{\rm o}^{\rm ini}\right)^2 - \sigma_\ast^2$ to estimate the size
of the pre-seeing galaxy image as the initial guess, where
$\sigma_\ast^2$ is the harmonic mean of the PSF sizes over different
exposures.  Note that, similarly to the PSF fitting, we account for the
astrometric distortion by performing the fitting in the celestial
coordinates.

Then, by using the coefficients $b^\ast_{pq}$ obtained from the PSF
estimation in Section~\ref{sec:shape_measurement_formalism}, we estimate
the ellipticity of the pre-seeing galaxy image for each galaxy by
minimizing
\begin{flalign}
&\chi^2 = \sum_{\eta=1}^{N_\mathrm{exp}} \sum_{\alpha=1}^{N_\mathrm
{pix}^{(\eta)}} &\nonumber\\
&\frac{\left[f^{(\eta)}_s 
I^{(\eta)}\hspace{-0.3em}\left(\bmf{\theta}^{(\eta)}_{\alpha}\right) -
\sum_{p,q} b_{pq} \phi_{pq}^{\sigma_{\rm o} E}\hspace{-0.3em}\left(\bmf{b}^{\ast
 (\eta)};
\bmf{{\cal
W}}^{(\eta)}\!\!\left(\bmf{\theta}^{(\eta)}_{\alpha}\right)\right)\right]^2}{\left(f^{(\eta)}_{s} \sigma^{(\eta)}_{\alpha}\right)^2},&
\end{flalign}
where $\phi_{pq}^{\sigma_{\rm o} E}\hspace{-0.3em}\left(\bmf{b}^{\ast
(\eta)}; \bmf{\theta}\right)$ are the basis functions including the PSF
convolution effect, defined as
\begin{equation}
\phi_{pq}^{\sigma_{\rm o} E}\hspace{-0.3em}\left(\bmf{b}^{\ast}; \bmf{\theta}\right) =
 \left[ \psi_{pq}^{\hat{\sigma}_{\rm gal}E} \otimes \sum_{p^\ast, q^\ast}
  b_{p^\ast q^\ast}^{\ast}\psi_{p^\ast q^\ast}^{\sigma_{\ast}} \right](\bmf{\theta}).
\label{eq:galaxy_basis_function}
\end{equation}
The convolution in the above equation can be done analytically.
Following \cite{Nakajima:2006} and using the initial guess of the galaxy
size $\sigma_{\rm gal}$ obtained from Eq.~(\ref{eq:chisq_native_fit}),
we do not vary the dilution parameter $\mu$ and fix the galaxy size
parameter $\hat{\sigma}_{\rm gal}$ in the above equation to
$\hat{\sigma}_{\rm gal}^2 = \sigma_{\rm gal}^2 + (f_p-1) \sigma_\ast^2$,
where we set $f_p=1.2$. Thus we used the slightly widened size parameter
than expected from the initial guess, $\sigma_{\rm gal}$, because in
\cite{Nakajima:2006} it is shown that this choice results in more
sensitive measure of the input shear in image simulations. Again,
by imposing the ``null test'' conditions, we minimize the above $\chi^2$
in order to estimate the ellipticity parameter $\bmf{g}$ for the galaxy,
which is used for weak lensing shear estimation.

Using the best-fit $\bmf{b}$ coefficients, we can estimate the total
signal-to-noise ratio or significance for measuring the flux of each
galaxy image \citep{Bernstein:2002}:
\begin{eqnarray}
\nu &=& \frac{f}{\sqrt{{\rm Var}(f)}}, 
\end{eqnarray}
where the flux is defined in terms of the coefficients $b_{pq}$ as
$f\equiv \sum_p b_{pp}$. The variances or uncertainties of the
coefficient, ${\rm Var}(f)$, can be properly estimated by propagating
the sky noise $\sigma_\alpha$ into the parameter estimation. We set the
order of Gauss-Laguerre function for galaxy fitting to 2 and that for
PSF fitting to 8, in order that the fit will converge even for noisy
images. The shear recovering accuracy test for this set up is described
below.

Using image simulations, we have tested the robustness of our shape
measurement. To be more precise, we used the elliptical exponential
profile for a model galaxy image. For modelling a star image, we used
double Gaussian functions:
\begin{eqnarray}
I^{\ast}(r;\sigma, f_I, f_\sigma)&\equiv&G(r;\sigma) + f_I G(r;f_\sigma \sigma) \\
G(r;\sigma)&\equiv&e^{-\frac{r^2}{2\sigma^2}},
\end{eqnarray}
where $G(r;\sigma)$ is an unnormalized Gaussian profile with width
$\sigma$, and we used $\sigma=0.75^{\prime\prime}/2\sqrt{2\log(2)}$
corresponding to $0.75^{\prime\prime}$ in FWHM and $(f_I,f_\sigma) =
(0.1,2.0)$. We also included Gaussian noise in the simulated images, as
a model of the sky noise. We studied the accuracy to which we can
recover the input weak lensing shear as a function of the flux $S/N$ and
size of simulated galaxy images.  We have found that, in order to have a
relative accuracy of shear better than 10 per cent,
$|\delta\gamma/\gamma|\le 0.1$, we need to use galaxies satisfying
$\nu>20$ and $\sigma_{\rm gal}>1.2$ pixels. The final number density
becomes 3.2~arcmin$^{-2}$. Hence, in the following weak lensing
analysis, we further impose these conditions for galaxy selection, and
will come back to this issue to discuss how the shear recovery accuracy
will affect the mass estimate in
Section~\ref{sec:results_systematics_measurement}.

\subsubsection{Residual Correlation}
\label{sec:residual_PSF}

\begin{figure}
\includegraphics[width=8cm]{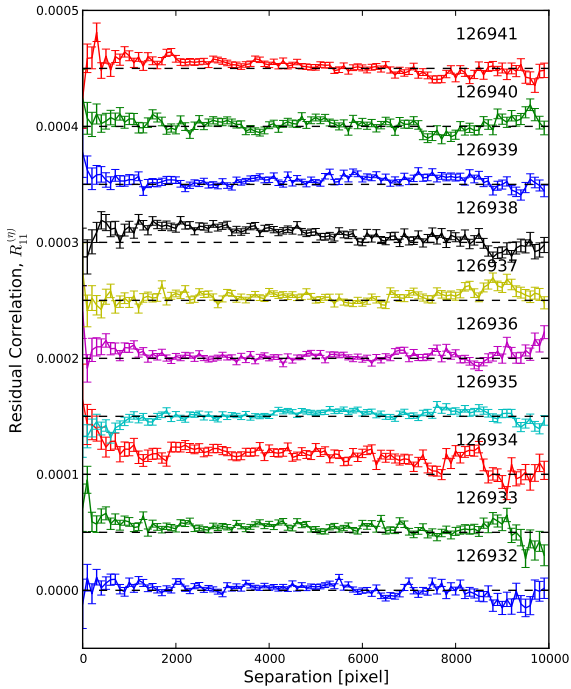} 
\caption{The residual correlation function (Eq.~\ref{eq:residual})
between the tangential components of the star ellipticities and the
galaxy ellipticities, against separation angle between the star and
galaxy pair.  Note that the tangential shear components are defined with
respect to the vector connecting star and galaxy in each pair, not with
respect to the cluster center.  For the galaxy ellipticity, we used the
difference between the galaxy ellipticities measured by combining all
the 10 exposures or the 9 exposures removing a particular one exposure
denoted by the label ID (e.g., 126932 for the first exposure). Hence the
data with error bars show the 10 different correlation functions. For
illustrative clarity, the functions except for the first exposure 126932
are vertically shifted (stepped by $5.0\times 10^{-5}$ for each curve),
and the dashed line around each result denotes the zero amplitude.}
\label{fig:residual_correlation}
\end{figure}

One of the great advantages of the multi-exposure fitting is that we
keep the PSF information in each exposure. In this section, we study
diagnostics for identifying an exposure that may not be suitable for
shape measurement, either in terms of data quality or inaccuracy of
PSF estimation, e.g. due to too rapidly-varying PSF patterns that cannot
be handled by the chosen PSF modeling algorithm. For this purpose, we
consider the following correlation function between the ellipticities of
galaxies and stars:
\begin{flalign}
R_{ij}^{(\eta)}(\theta) \equiv &
\left\langle e_i^{{\rm star},(\eta)}(\bmf{\theta}^{\prime})
\right. \nonumber\\ 
&\hspace{-1em}\times \left. \left(e_j^{{\rm gal}, ({\rm all})}(\bmf{\theta}'+\bmf{\theta}) 
- e_j^{{\rm gal},({\rm all}-\eta)}(\bmf{\theta}'+\bmf{\theta})\right)
\right\rangle,
\label{eq:residual}
&
\end{flalign}
where $\langle\cdots \rangle$ denotes the average for all the pairs
separated by the angle $\theta$; 
$e_i^{{\rm star},(\eta)}(\bmf{\theta}')$ is the $i$-th ellipticity
component of star at the position $\bmf{\theta}'$ for the $\eta$-th
exposure; $e_i^{{\rm gal}, ({\rm all})}(\bmf{\theta}'+\bmf{\theta})$ is
the ellipticity component of galaxy at the position
$\bmf{\theta}'+\bmf{\theta}$, measured by combining all the exposures;
$e_i^{{\rm gal}, ({\rm all}-\eta)}(\bmf{\theta}'+\bmf{\theta})$ is the
ellipticity measured by combining the exposures 
except for
the $\eta$-th
exposure.  
Although the correlation between star and galaxy ellipticities is often
used in the literature as a diagnostic of the imperfect shape
measurement, 
the above correlation can be more useful for identifying problems with
some 
particular exposure, as explained below.

Suppose that the $\eta$-th exposure has a systematic error in the PSF
estimation. In this case, $e_i^{{\rm gal}, ({\rm all})}(\bmf{\theta})$
may have some contamination from the imperfect PSF estimation in the
$\eta$-th exposure, while $e_i^{{\rm gal} ({\rm
all}-\eta)}(\bmf{\theta})$ does not have the contamination.  The
difference $[ e_i^{{\rm gal},({\rm all})}(\bmf{\theta}) - e_i^{{\rm
gal},({\rm all}-\eta)}(\bmf{\theta}) ]$ is sensitive only to the PSF
estimation of the $\eta$-th exposure. Hence, if the imperfect PSF
estimation is really a problem, the ellipticity difference may have a
non-vanishing correlation with the PSF ellipticity of the $\eta$-th
exposure, $e_i^{{\rm star},(\eta)}(\bmf{\theta})$. This is what the
correlation (Eq.~\ref{eq:residual}) tries to measure. Hereafter we call
this the {\em residual correlation}.  Its advantage over a direct
correlation is that, in such small fields as these, the PSF ellipticity
can easily correlate with the real lensing shear; such an effect cancels
out of the difference in the residual correlation, but would contribute
to a standard star-galaxy correlation function.

Since we have 10 different exposures for the $i'$-band image of the
ACTJ0022 data, we have 10 different correlation functions to test the
accuracy of PSF estimation in each exposure.
Fig.~\ref{fig:residual_correlation} shows the results. The figure
clearly shows that one exposure with ID ``126934'' shows non-zero
correlations over all the range of separation angles, indicating that
the exposure has some systematic issue in the PSF estimation.  In fact,
we found that the PSF in this exposure exhibits larger ellipticities,
typically $e\sim 0.04$, than in other exposures.  Although we have
checked that the weak lensing tangential shear signal is not
significantly affected even when including the exposure in the analysis,
we do not use the 126934 exposure in the following analysis\footnote{One
might be concerned that the nonzero residual correlation suggests that
we should not trust the PSF size estimate, which could give rise to a
multiplicative bias in the shear.}.  One may notice that the other
residual correlations show non-vanishing correlations with amplitude
$\sim 10^{-5}$ at some scales. Since the ellipticity difference, $[
e_i^{{\rm gal},({\rm all})}(\bmf{\theta}) - e_i^{{\rm gal},({\rm
all}-\eta)}(\bmf{\theta}) ]$, arises naively from the star ellipticities
in the $\eta$-th exposure, the residual correlation would scale as
$(e^{{\rm star}, (\eta)})^2$. In turn, if the galaxy ellipticity is
affected by the imperfect PSF correction inferred by the residual
correlations, the contamination to the cluster lensing would be of the
order of $e^{{\rm star}, (\eta)}\simeq \sqrt{R}\sim 0.003$, which is
more than one order magnitude smaller than the cluster lensing. Hence we
do not believe that the residual PSF systematic error, even if it
exists, should affect the following weak lensing analysis (see later for
further discussion on the impact of imperfect PSF estimation).

\section{Results}
\label{sec:results}
\subsection{Cluster Mass}
\label{sec:results_mass}

We can now combine the photo-$z$ estimate and shape measurement for each
background galaxy to estimate the weak lensing signal of ACTJ0022. In
this paper, we focus on the tangential shear component, defined as
\begin{equation}
g_+ = -\left(g_1\cos2{\phi} + g_2\sin2{\phi}\right),
\end{equation}
where $\phi$ is the position angle between the $1$st coordinate axis and
the vector connecting the galaxy position and the cluster center for
which we use the BCG position. Similarly, we can define the component,
$g_\times$, from the 45 degrees rotated ellipticity component from $g_+$.

To estimate the weak lensing signal due to ACTJ0022, we compute the
radial profile by averaging the measured tangential ellipticities of
background galaxies in each circular annulus as a function of the
cluster-centric radius:
\begin{eqnarray}
\left\langle e_+(\theta_n) \right\rangle &=& \frac{1}{{\cal R}}
 \frac{\sum_i w_i e_{+,i}}{\sum_i w_i}, 
\end{eqnarray}
where $w_i$ is the weight for the $i$-th galaxy, the summation $\sum_i$
runs over all the galaxies lying in the $n$-th annulus with radii
$\theta_{n,{\rm in}}\le \theta\le \theta_{n,{\rm out}}$, and ${\cal R}$
is the shear responsivity.  To compute $w_i$ and ${\cal R}$, we used
Eqs.~(5.33), (5.35) and (5.36) in \cite{Bernstein:2002}.  Note that, for
the central value of each radial bin, we infer the area-weighted mean
radius of the annulus, i.e. $\theta_n\equiv \int_{\theta_{n, {\rm
in}}}^{\theta_{n, {\rm out}}}2 \pi r^2\mathrm{d}r / \int_{\theta_{n,
{\rm in}}}^{\theta_{n, {\rm out}}}2 \pi r \,\mathrm{d}r$. Similarly we
estimate the statistical uncertainty of the measured signal in each
radial bin:
\begin{eqnarray}
\sigma_{e_+}(\theta_n) &=& \frac{1}{{\cal R}}\sqrt{
 \frac{\sum_i w_i^2 e_{+,i}^2}{\left(\sum_i w_i\right)^2}}.
\end{eqnarray}
Here we have assumed that the statistical uncertainty arises solely from
the intrinsic ellipticities of source galaxies per component.  Recalling
that the relation between the ellipticity ($e$) and the shear ($g$) is
given as $e=\tanh(2\tanh^{-1}g)$, where $e=\sqrt{e_+^2+e_\times^2}$ and
so on, we can convert the measured ellipticities to the lensing shear
components; e.g., $g_+=(g/e)e_+$.

\begin{figure*}
\includegraphics[width=8cm]{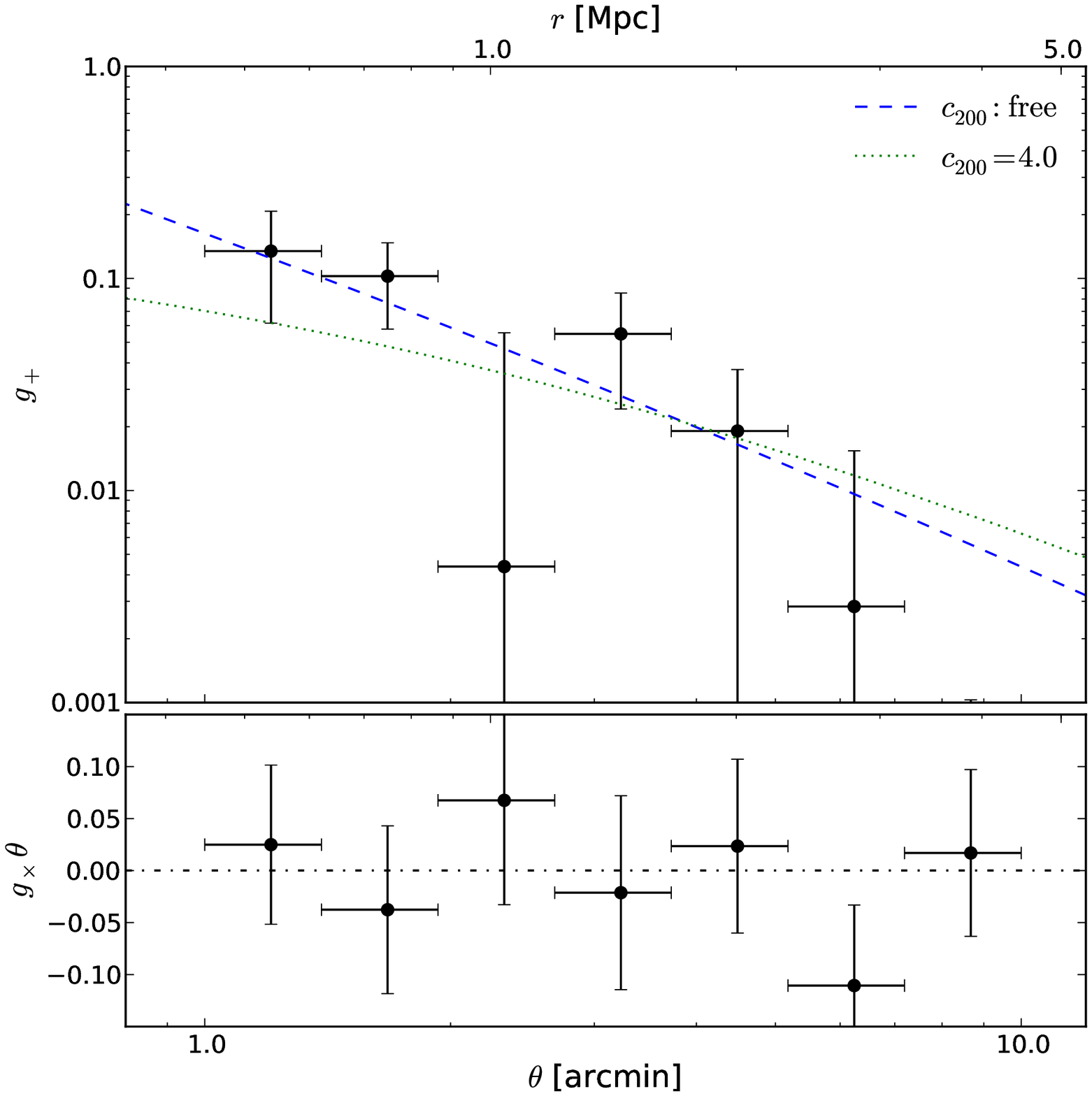}
\includegraphics[width=8cm]{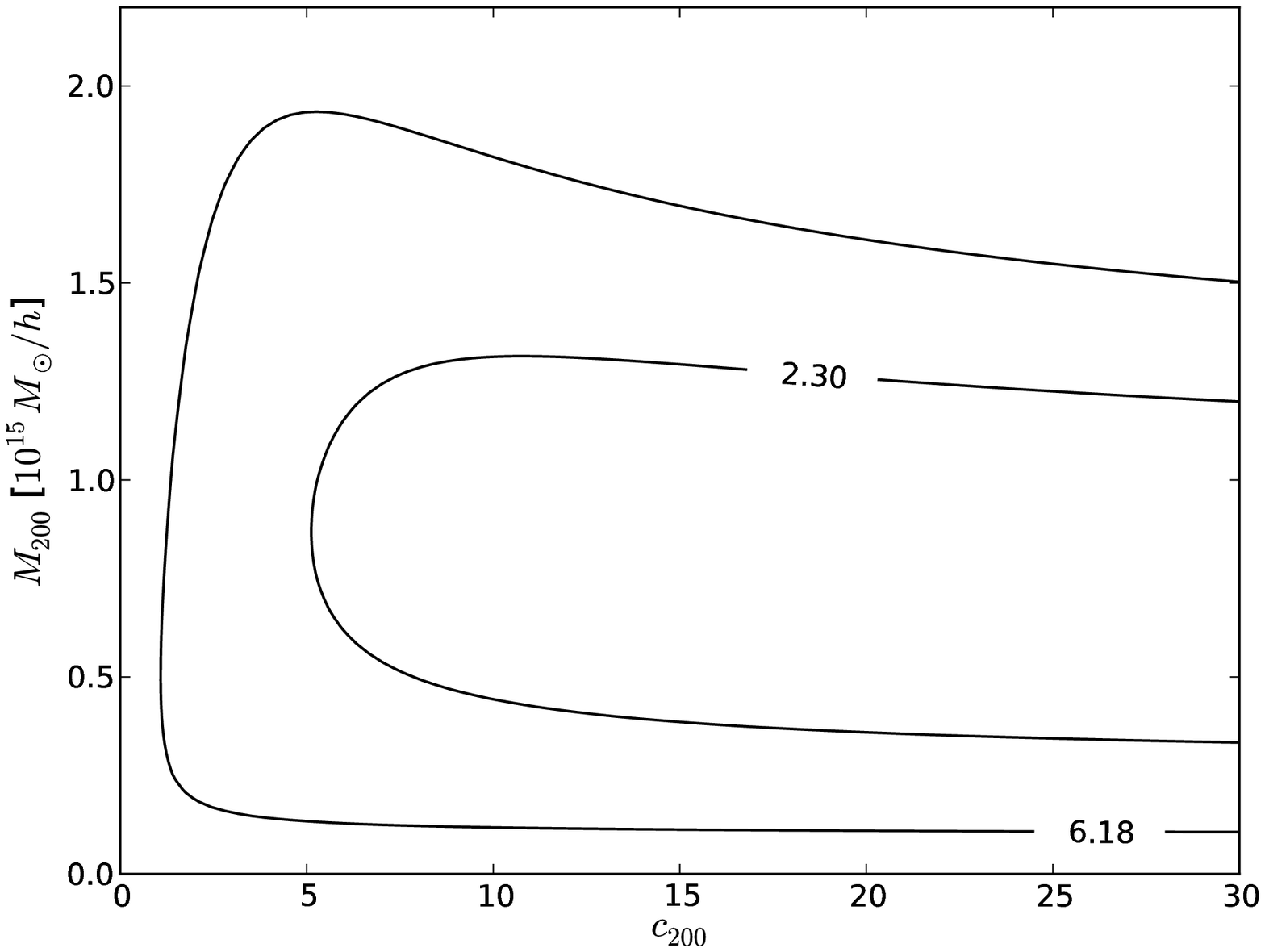} 
\caption{{\em Left panel:} The measured radial profiles of tangential
shear component (upper plot) and its 45-degrees rotated component,
non-lensing mode (lower plot).  The vertical error bar around each data
point shows the $1\sigma$ statistical error in each radial bin, while
the horizontal error bar denotes the bin width. The dashed curve shows
the best-fit NFW profile, while the dotted curve is the best-fit NFW
when fixing the concentration parameter to the $\Lambda$CDM model
expectation, $c_{200}=4.0$ (see text for details). The non-lensing
B-mode, $g_\times$, is consistent with zero over a range of the radial
bins we consider.  {\em Right panel:} The $\Delta \chi^2$ contours in
($M_{200}, c_{200}$) plane for the NFW profile fitting where the
concentration parameter is allowed to vary. The two lines correspond to
$\Delta \chi^2 = 2.30$ (68\% C.L.) and $6.18$ (95\% C.L.),
respectively.}

\label{fig:shear}
\end{figure*}

Fig.~\ref{fig:shear} shows the measured radial profiles for the
tangential shear and the 45-degree rotated component for ACTJ0022.  The
figure clearly shows the coherent signals for $g_+$, where the
amplitudes are increasing with decreasing radius as expected for cluster
lensing. On the other hand, the non-lensing mode $g_\times$, which can
serve as a monitor of the residual systematic effects, is consistent
with zero over the range of radii we consider. Note that we plot the
$g_\times$-profile in units of $\theta_n \times g_\times(\theta_n)$ so
that the scatter in the values is independent of radius for
logarithmically-spaced binning, if the measurement errors in the
$g_+/g_\times$ signals arise from the random intrinsic
shapes\footnote{The number of background galaxies in each annulus scales
with radius as $N_g\propto \theta_n^2\Delta \ln\theta$ for the
logarithmically-spaced binning. The shape noise contribution to the
statistical errors of the $g_+/g_\times$ measurements scale as
$\sigma(g_{+,\times})\propto \sigma_\epsilon^2 /\sqrt{N}_g\propto
1/\theta_n$. Hence $\theta_n \sigma(g_{+,\times})$ becomes independent
of radius.}.  However, the shear measurement is still noisy, mainly due
to the small number density of source galaxies (3.2 arcmin$^{-2}$).  If
we estimate the total $S/N$ for the shear measurement as $(S/N)^2\equiv
\sum_n \left[\langle g_+(\theta_n)\rangle/ \sigma_+^2(\theta_n)\right]$,
we find $S/N\simeq 3.7$, i.e. about 3.7$\sigma $ detection of the
lensing signal.

We now estimate the cluster mass of ACTJ0022 by comparing the measured
shear signal to the model lensing profile expected from the
Navarro-Frenk-White (NFW) profile \citep{Navarro:1996}. The NFW profile
is given as $\rho_{\rm NFW}=\rho_s/[(r/r_s)(1+r/r_s)^2]$ and specified
by two parameters ($\rho_s, r_s$). We can rewrite the NFW profile to be
specified by the enclosed mass $M_\Delta$ and the concentration
parameter $c_\Delta$ \citep[e.g. see][for the conversion]{Okabe:2010}.
The cluster mass often used in the literature is the three-dimensional
mass enclosed within a spherical region of a given radius $r_\Delta$
inside of which the mean interior density is $\Delta$ times the mean
mass density at the cluster redshift, $\bar{\rho}_m(z_l)$:
\begin{equation}
M_\Delta = \frac{4\pi}{3} r_\Delta^3 \bar{\rho}_m(z_l) \Delta.
\label{eq:enclosed_mass}
\end{equation} 
Note that, in this analysis, we are working in physical distance units.
The concentration parameter is defined by $c_\Delta =r_\Delta /r_s$. For
most of this paper, we use $\Delta=200$.  Alternatively, the cluster
mass can be defined in terms of the critical density $\rho_c$ instead of
$\bar{\rho}_m$, in which case we denote the mass as $M_{\Delta \rho_c}$
in the following.

Given the NFW profile, we can analytically compute the expected radial
profiles of the lensing fields \citep{Bartelmann:1996,Wright:1999}. For
example, the lensing convergence profile, which is equivalent to the
radial profile of the projected mass density, is computed as
\begin{equation}
\kappa_{\rm NFW}(\theta)\equiv \Sigma_{\rm
 cr}^{-1}\int_{-\infty}^{\infty}
\!dr_\parallel~
\rho_{\rm NFW}\left(\sqrt{r_\parallel^2+(D_l\theta)^2}\right),
\end{equation}
where $\Sigma_{\rm cr}$ is the critical surface mass density (see below)
and $D_l$ is the angular diameter distance to the cluster redshift. The
projection integration in the above equation can be analytically
done. Similarly the shear profile $\gamma_{\rm NFW}(\theta)$ can be
analytically derived. The measured shear profile $g_+(\theta)$ is the
reduced shear \citep{BartelmannSchneider:2001}, and is given as
$g_+(\theta)=\gamma_{\rm NFW}(\theta)/(1-\kappa_{\rm NFW}(\theta))$ for
an NFW profile.  The critical surface mass density is given as
\begin{equation}
\Sigma_{\rm cr} = \frac{c^2}{4 \pi G} D_l^{-1} \left\langle
						\frac{D_{ls}}{D_s}\right\rangle^{-1},
\label{eq:sigma_cr}
\end{equation}
where $D_l$, $D_s$, and $D_{ls}$ are angular diameter distances from
observer to cluster (lens), from observer to source, and from cluster to
source. The mean distance ratio is calculated using the photo-$z$
estimates of source galaxies as
\begin{equation}
R \equiv \left\langle \frac{D_{ls}}{D_s} \right\rangle =  \frac{\sum_i
 w_i \left[1-D_l/D(z_{{\rm phz},i}) \right]}{ \sum_i w_i},
\label{eq:mean_distance_ratio}
\end{equation}
where the summation runs over all the source galaxies and $w_i$ is the
weight used when calculating the shear profile.  Note the average above
is equivalent to the average $\langle 1/D(z_s)\rangle$, as the cluster
redshift (lens redshift) is known.

\begin{table}
 \begin{center}
  \begin{tabular}{@{}ccccc@{}}
  \hline \hline
   & setup& $M_{200}\left[10^{15}M_{\odot}/h\right]$ & $c_{200}$ &
   $\chi^2/{\rm d.o.f.}$\\
 \hline
   Case 1 & $c_{200}$: free & $0.75^{+0.32}_{-0.28}$ & $>9.7$ & $4.38/5$ \\
   Case 2 & $c_{200}=4.0$ & $0.85^{+0.55}_{-0.44}$ & fixed & $7.29/6$ \\
\hline
\end{tabular}
  \caption{Results for the NFW profile fitting to the measured
  tangential shear profile for ACTJ0022 shown in Fig.~\ref{fig:shear}.}
\label{tab:nfwfit}
\end{center}
\end{table}

We estimate the cluster mass $M_\Delta$ by minimizing the following
$\chi^2$ with varying the model parameters ($M_\Delta, c_\Delta$):
\begin{equation}
\chi^2 = \sum_n \frac{\left[ \left \langle g_+(\theta_n) \right \rangle
		       - g_{\rm
      NFW}(\theta_n; M_\Delta, c_\Delta)\right]^2}{\sigma_{g+}(\theta_n)^2}.
\label{eq:nfw_chisquare}
\end{equation}
We consider two cases for the NFW fitting: For Case 1, we allow the
concentration parameter to be free; for Case 2, we fixed it to $c_{200}
= 4.0$, which is a theoretically-expected $1\sigma$ upper bound on the
concentration parameter for a cluster with $M_{200}=10^{15} M_\odot/h$.
To be more precise, the fitting formula derived in \cite{Duffy:2008}
using N-body simulations for a $\Lambda$CDM model gives $c_{200}\simeq
3.2$ for a cluster with $M_{200}=10^{15}M_\odot/h$ and at $z=0.81$.
Since the Subaru WL prefers a steeper NFW profile (therefore with the
higher $c_{200}$) for ACTJ0022 as we will discuss below, we adopt the
$1\sigma$ upper bound of $c_{200}=4.0$ motivated by the fact that the
simulations show typical intrinsic scatters of $\sigma(c_{200})\simeq 1$
for such massive halos.

Table~\ref{tab:nfwfit} shows the results for the two cases, and the left
panel of Fig.~\ref{fig:shear} shows the best-fit NFW profiles compared
with the measurement. For Case 1, we cannot constrain the concentration
parameter, and obtained only the $1\sigma$ lower bound as $c_{200}\ge
9.7$, because the measured shear profile does not show a clear curvature
over the range of radii we probe. The lower bound also means that the
measured shear profile is consistent with the outer part of NFW profile,
$\rho_{\rm NFW}\propto r^{-3}$. This can be explained as follows. The
best-fit virial radius $r_{200}\simeq 1.8~$Mpc indicates the NFW scale
radius $r_s\sim 0.5~$Mpc if we assume the concentration parameter $c\sim
4$, the $\Lambda$CDM prediction. As shown in Fig.~\ref{fig:shear}, the
shear signals at radii smaller than 0.5~Mpc are not available, meaning
that we cannot probe the inner part of the expected NFW profile from the
measured shear signal and constrain the concentration parameter from the
varying slope of the profile. If strong lensing signals are available
for the inner regions, we may be able to constrain the concentration
parameter as done in \cite{Broadhurst:2005}, but we have not found any
strongly-lensed candidates in the cluster region.

For Case 2, we found a slightly larger best-fit mass than in Case 1,
because the concentration parameter is fixed to $c_{200}=4.0$, which is
smaller than the 1$\sigma$ lower bound for Case 1, and a larger mass is
needed to explain the measured shear amplitude with the small $c_{200}$
(see the right panel of Fig.~\ref{fig:shear}).  However, the difference
between the best-fit cluster masses for Case 1 and 2 is within the error
bars, so not significant.

\subsection{Systematic Uncertainties from Measurement}
\label{sec:results_systematics_measurement}

In this section, we discuss the impact of several systematic errors
on the cluster mass estimation.

\subsubsection{Imperfect shape measurement}

First, we consider systematic error due to imperfect shape measurement.
To estimate the impact, as described in
Section~\ref{sec:shape_measurement_selection_criteria}, we have carried
out many image simulations as a function of different flux $S/N$ values
and the different galaxy size parameters for simulated galaxy images. We
considered $\nu=20, 27,60,130$ and $\sigma_{\rm gal}=1.3, 1.4, 1.8, 2.2,
2.7$ pixels, in total 20 different image simulations. For each
simulation that contains 80000 galaxies, we tested whether our shear
method can recover the input shear. For each simulation, we quantify the
systematic error found from the image simulations in terms of a
multiplicative bias parameter $m$: $\gamma_{\rm
recovered}=(1+m)\gamma_{\rm input}$. We have found that our method leads
to a 1 per cent to 10 per cent bias, or $m=0.01-0.1$, where $m$ is
determined within relative accuracy of $\sim$10 per cent.  Then, we
averaged the simulation results for the estimated bias by weighting the
result of each simulation with the number density of galaxies used for
our actual ACTJ0022 analysis that fall into the similar region of the
flux $S/N$ and size values of each simulation.  As a result, we found
the average multiplicative bias $m\simeq -0.06$ for the background
galaxies of ACTJ0022, implying that our method tends to underestimate
the true shear value and therefore underestimate the cluster mass.

\subsubsection{Photo-$z$ errors}

We study how photo-$z$ errors used in selecting background galaxies
affect the cluster mass estimate. There are two effects to be
considered: (1) a dilution of the lensing signals caused by an inclusion
of unlensed galaxies into the background galaxy sample, and (2)
inaccuracy in estimating the mean critical mass density $\Sigma_{\rm
cr}$ from the photo-$z$s (Eq.~\ref{eq:sigma_cr}).

For the dilution effect, the correction factor is estimated from the
fraction of galaxies whose true redshifts are lower than the cluster
redshift $0.81$:
\begin{equation}
f_c \equiv \frac{N_{{\rm sel}, z_p}(z_s < 0.81)}{N_{{\rm sel}, z_p}},
\end{equation}
where $z_s$ is the true redshift, and $N_{{\rm sel}, z_p}$ is the total
number of galaxies in the background galaxy catalog. We checked that the
radial profile of number densities of the background galaxies does not
show any radial dependence, i.e. no clear indication of the
contamination of unlensed cluster member galaxies.  Nevertheless, we
here address an effect of possible residual contamination from
foreground galaxies on the lensing signal.  If the contamination is
uniform over the ACTJ0022 field, as indicated by the number density
profile, the measured shear is diluted as
\begin{equation}
\left \langle g^{\rm meas} \right \rangle = (1-f_c)\left \langle g^{\rm true} \right \rangle,
\end{equation}
where $\langle g^{\rm meas}\rangle$ and $ \langle g^{\rm true}\rangle$
and the measured and underlying-true shear signals, respectively. If
$f_c>0$, the measured shear signal is affected by the dilution, and
therefore underestimated. The true shear and  the true cluster
mass should be higher than inferred from the measurement.

For inaccuracy in the $\Sigma_{\rm cr}$ estimation, the correction
factor can be estimated as
\begin{equation}
R^{\rm true} \equiv
\frac{\displaystyle
\int_{z_{\rm lens}}^\infty
\mathrm{d}z_s 
\frac{\mathrm{d}N_{{\rm sel},z_p}^{\rm
		     true}}{\mathrm{d}z_s}
\frac{D_{ls}(z_s)}{D_s(z_s)} 
}
{\displaystyle 
\int_{z_{\rm lens}}^\infty \mathrm{d}z_s
		     \frac{\mathrm{d}N_{{\rm sel},z_p}^{\rm
		     true}}{\mathrm{d}z_s}},
\end{equation}
where $\mathrm{d}N^{\rm true}/\mathrm{d}z_s$ is the underlying true
redshift distribution of the background galaxies. The question is
whether the quantity $R$, estimated based on the photo-$z$s
(Eq.~\ref{eq:mean_distance_ratio}), may differ from the true value
$R^{\rm true}$ due to the photo-$z$ errors. If there is a bias in $R$,
denoted as $R = R^{\rm true} + \delta R$, the NFW profile to be compared
with the measured shear profile is biased as
\begin{equation}
g_{\rm NFW}^{\rm phz} \equiv g_{\rm NFW}^{\rm true}\left( 1 + \delta R /R^{\rm true}\right),
\end{equation}
where $g^{\rm phz}_{\rm NFW}$ is the model NFW inferred from the
photo-$z$ information of every galaxy and $g_{\rm NFW}^{\rm true}$ is
the model NFW profile using the true distance ratio. 
If $\delta R>0$, the model NFW amplitude is overestimated, and then
the best-fit mass would be underestimated in order to reproduce the
measure shear amplitude. Hence the true mass should be higher than
inferred.

To estimate possible biases in the factors $f_c$ and $R$, we used the
publicly available COSMOS photo-$z$ catalog assuming that the photo-$z$s
derived by using 30 broad, intermediate and narrow-band data are true
redshifts \citep{Ilbert:2009}. We obtain the photo-$z$ distribution for
the COSMOS galaxies by applying our photo-$z$ method to the COSMOS
$Br'i'z'$ magnitudes of each galaxy to estimate its photo-$z$. Note that
the COSMOS catalog does not have the $Y$-band data.  Since the limiting
magnitude of the background galaxies used for the weak lensing analysis
($i'_{\rm lim}=25.6 $) is shallower than the COSMOS catalog ($i'_{\rm
lim}=26$), we can reliably use the COSMOS catalog for this purpose. To
correct for the limiting magnitude difference, we use the following
equation to estimate the underlying true redshift distribution for the
background galaxy sample:
\begin{align}
  \label{eq:cosmos_nz_scale}
  \frac{dN_{\rm sel, zp}^{\rm ACTJ}}{dz_s}
  =
  \frac{dN_{\rm sel, zp}^{\rm COSMOS}}{dz_s} \times
  \frac{dN^{\rm th}/dz(i<25.6)}{dN^{\rm th}/dz(i<26)}
\end{align}
where $dN^{\rm th}/dz$ is the fitting formula that gives the
redshift distributions as a function of the limiting magnitude in
\cite{Ilbert:2009}. Using the redshift distribution given by
Eq.~(\ref{eq:cosmos_nz_scale}), we found that possible biases in the
correction factors are $f_c\simeq 0.10$ or $\delta R/R^{\rm true}\simeq
- 0.07$, respectively.

The COSMOS photo-$z$ catalog may be affected by cosmic sample variance
due to the small area coverage (about 2 deg$^2$); the redshift
distribution shows non-smooth features due to large-scale structures
along the line-of-sight. Hence we also estimate the impact of photo-$z$
errors using the mock catalog used in \cite{Nishizawa:2010}.  In the
mock catalog, we properly included the response functions of Subaru
$Br'i'z'Y$ filters we used.  We generated the mock catalog such that it
reproduces the fitting formula for the redshift distribution of the
COSMOS photo-$z$ catalogs in \cite{Ilbert:2009} as a function of the
$i'$-band limiting magnitudes.  Note that the fitting formula for the
redshift distribution has a smooth functional form against redshift.  We
also included a mixture of different galaxy SED types according to the
COSMOS results.  By estimating photo-$z$s for the mock galaxies and
using galaxies down to the limiting magnitude of ACTJ0022, we found
biases of 0.15 for $f_c$ and 0.07 for $\delta R/R^{\rm true}$,
respectively.

From the above investigation, we estimate typical bias from inaccurate
photo-$z$ estimation to be $f_c\simeq0.10$ and $ \delta R/R^{\rm true}
\simeq \pm 0.07$. Note that we further imposed size and magnitude cuts
on the background galaxies for the weak lensing analysis, which
preferentially selects brighter galaxies than the limiting
magnitude. Hence, the biases inferred here correspond to a maximum bias,
because the brighter galaxies have more accurate photo-$z$ and are less
contaminated by photo-$z$ outliers.

\subsubsection{Imperfect PSF estimation}

Although we carefully tested for imperfect PSF estimation in
Section~\ref{sec:residual_PSF}, here we consider how a residual
systematic error in the PSF estimation affects the shear estimation.

First, we consider the impact of the PSF size misestimation as studied
in the left panel of Fig.~\ref{fig:ellipticity_calibrate}, where we
found a 0.2 per cent level in the size misestimation. Following the
prescription provided by \cite{Hirata:2004}, we found that the PSF size
misestimation of 0.2 per cent corresponds to typically 0.2 per cent in
the shear bias, so this is negligible compared to the statistical error.

Second, we consider the effect caused by a misestimation of the PSF
ellipticities. One nice feature of the cluster lensing measurement is
that it measures the coherent tangential shear pattern inherent in
background galaxy ellipticities with respect to the cluster center, but
an imperfect estimate of PSF ellipticities may not necessarily mimic the
tangential shear pattern. As a possible maximum effect, assuming
completely ineffective PSF correction for very poorly-resolved galaxies,
we simply calculated the average of star ellipticities in each annular
bin used for the shear analysis. The average ellipticity is consistent
with zero in the outer radial within the standard deviation, but the
average in the second and third bins deviates from zero by more than
$2\sigma$: $\langle{g^\ast}\rangle \simeq -0.006$. We can estimate a
maximum effect by assuming that the average PSF ellipticity propagates
into the systematic error of the shear estimate, which should not be the
case after the PSF correction. The bias $\delta g_+/g_+\simeq
-0.006/0.1\simeq -0.06$, is $-6$ per cent, where $g_+\simeq 0.1$ is the
shear amplitude in the inner bins as shown in Fig.~\ref{fig:shear}.

\subsubsection{Total budget of systematic errors on cluster mass}

We can now sum up all the systematic errors in the shear estimates we
have so far described. If a shear bias is negative, such that $m<0$
for the shear multiplicative bias, the true shear value should be
higher, and in turn the true cluster mass is higher than estimated.
Thus we refer to possible corrections in the cluster mass, according to
the systematic errors of the cluster mass; e.g., for the multiplicative
shear error of $-6$ per cent ($m=-0.06$), we refer to the correction in
the cluster mass as ``+7 per cent''. We found such a possible correction
in the cluster mass by re-fitting the NFW profile to reduced shear which
is manually corrected for the bias predicted in the previous
subsections.

Summing up these possible systematic errors in quadrature, the total
amount of the correction in the cluster mass is estimated as +17 per
cent and $-$8 per cent, which is about half of the statistical error in
Table~\ref{tab:nfwfit}.

\subsection{Systematic Uncertainties from Physical Considerations}
\label{sec:results_systematis_physics} In
Section~\ref{sec:results_mass}, we constrained the cluster mass by
deprojecting two-dimensional lensing information assuming that the mass
distribution of ACTJ0022 follows a spherically-symmetric NFW
profile. However, dark matter halos are triaxial in general, as seen in
$\Lambda$CDM simulations \citep{Jing:2002}.  Thus the mass estimate
assuming spherical symmetry can be biased. \cite{Oguri:2005} estimated
the halo triaxiality effect on lensing measurements, and showed that the
mass can be biased by $\pm 20 $ -- $30$ per cent depending on the
projection direction.  The amount of the possible mass bias corresponds
to $\pm 50 $ -- $70$ per cent ($\pm 30$ -- $50$ per cent) of the
statistical error when the concentration parameter is free (fixed).

When fitting an NFW profile to the tangential shear profile, we assumed
that the BCG position (R.A.=00:22:13.04, Dec.=$-$00:36:33.84) is the
cluster center. However, the BCG may have an offset from the true center
of the dark matter halo hosting the cluster. If an off-centered BCG is
assumed to be at the center of the dark matter halo profile, the
tangential shear signal is diluted at radii smaller than the offset
radius \citep{Oguri:2011}.  By using the halo centers inferred by
various observables such as X-ray and/or distribution of satellite
galaxies for low-redshift clusters at $z\sim 0.2$, previous work showed
that a typical displacement, if it exists, is about 2--3 percent of the
virial radius\footnote{This statement is true in the absence of
photo-$z$ errors, which can cause selection of the wrong galaxy as the
BCG.} \citep{vandenBosch:2005, Koester:2007, Bildfell:2008}. Assuming a
similar amount of displacement for the BCG of ACTJ0022 at $z=0.81$, we
study how the shear signals are changed. To be more precise, we
re-calculated the tangential shear measurements by taking 8 different
centers along the circle of radius 0.03$r_{\rm vir}$, with different
position angles stepped by 45 degrees ($\theta=0, 45, \dots, 315$
degrees). Note that we employed the best-fit $r_{\rm vir}$ for Case 1
and 2, respectively, as given in Table~\ref{tab:nfwfit}.  We found that
7 (6) out of the 8 different centers yield smaller best-fit cluster
masses for Case 1 (2), respectively. Hence, a possible bias in the
cluster mass due to the offset is estimated as 10 (7) per cent for Case
1 (2).  This result implies that the BCG position is close to the true
center.  The BCG center is also supported by the high-resolution SZ
observation, done by \cite{Reese:2012} with SZA; the estimated center is
R.A.=00:22:13.006, Dec.=$-$00:36:33.35 with error of
$0.8^{\prime\prime}$ and $1.1^{\prime\prime}$, respectively, in good
agreement with the BCG position.

The mass distribution at different redshifts along the same
line-of-sight of ACTJ0022 may contaminate the weak lensing signal -- the
so-called projection effect. The projection effect is equivalent to weak
lensing due to large-scale structures (hereafter simply cosmic shear),
and acts as a statistical noise to the cluster lensing. It is difficult
to quantify the impact of the projection effect on individual cluster
lensing, unless a prominent structure at a different redshift is
identified, e.g., from a concentration of galaxies, which we do not
find. Here we estimate the systematic error by assuming the typical
projection effect expected for the $\Lambda$CDM model. We follow the
method in \cite{Oguri:2011} (see discussion around Eq.~47) in order to
include the covariance error matrix between the tangential shear signals
of different radii due to the typical projection effect for the
$\Lambda$CDM model. Then we re-did the $\chi^2$-fitting, and found that
the the best-fit cluster mass is changed only by about $+4$ per cent,
which is much smaller than the statistical error due to the shape
noise. The statistical error of mass estimate is increased by $+3$ per
cent due to the covariance error matrix.

\subsection{Cosmological Implications}
\label{sec:cosmological_implications}

\subsubsection{Scaling Relation}
\begin{figure}
\begin{center}
\includegraphics[width=9cm]{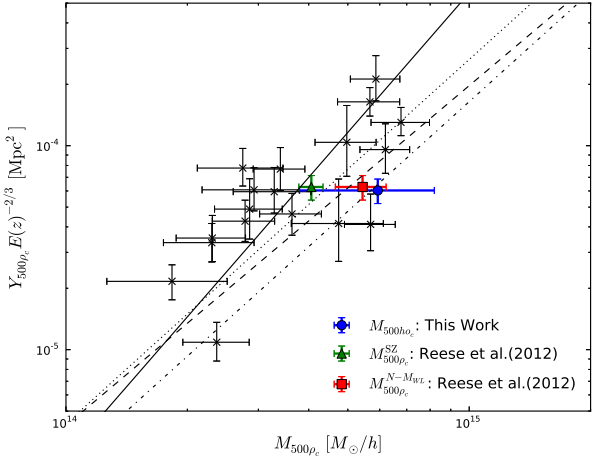}
\caption{ The circle symbol with error bars shows our weak lensing
results for ACTJ0022 ($z=0.81$), in the plane of the Compton-$y$
parameter $Y_{500\rho_c}$ and the cluster mass $M_{500\rho_c}$ (for the
overdensity of 500 times the critical density). Note that the
$Y_{500\rho_c}$ value quoted is taken from \protect\cite{Reese:2012} and
$E(z)$ is the redshift evolution of the Hubble expansion rate. For
comparison, the triangle and square symbols (slightly shifted vertically
for clarity) are taken from \protect\cite{Reese:2012}, showing the mass
estimates for ACTJ0022 derived using the SZA observation with the virial
theorem and the optical-richness and mass scaling relation of the SDSS
clusters, respectively. The star symbols are from
\protect\cite{Marrone:2011}, derived using the SZA observations and the
weak lensing mass estimates for the 18 LoCuSS clusters at redshift
$z\simeq 0.2$.  The solid line denotes the best-fit scaling relation.
The dashed line denotes the scaling relation in
\protect\cite{Andersson:2011} for the SPT SZ clusters in the wide
redshift range up to $z\sim 1$, which is derived combining the SZ
observation with the X-ray follow-up observations. The dotted line
denotes the scaling relation in \protect\cite{Arnaud:2010}, derived
using the X-ray observations for low-redshift clusters at $z\simlt
0.2$. The dot-dashed line denotes the scaling relation in
\protect\cite{sifon:2012}, derived using the dynamical mass estimates
for the ACT SZ clusters in the wide redshift range up to $z\sim1$.}
\label{fig:sz_m_scaling_ralation}
\end{center}
\end{figure}

The lensing mass of ACTJ0022 can be compared with the mass estimate in
\cite{Reese:2012}, where the two kinds of mass estimates were shown
using the deep SZA observation and the SDSS Stripe 82 data
\citep{Frieman:2008}.  First, they estimated the cluster mass from the
observed Compton-$y$ parameter assuming the hydrostatic equilibrium and
the universal pressure profile that is derived from the X-ray
observations of 33 low-redshift clusters ($z\simlt 0.2$) in
\cite{Arnaud:2010}. With the surface pressure correction proposed in
\cite{Mroczkowski:2011}, the cluster mass was estimated as
$M_{500\rho_c}^{SZ}=(0.40 \pm 0.03) \times 10^{15} M_\odot /h$.  Second,
they used the scaling relation between the optical richness (the number
of member galaxies) and the weak-lensing masses, done in
\cite{Rozo:2009} for the MaxBCG catalog \citep{Koester:2007} , in order
to infer the mass of ACTJ0022 assuming that the scaling relation holds
for the high redshift of ACTJ0022. Then the cluster mass was derived as
$M_{500\rho_c}^{N-M_{WL}} = (0.54\pm0.08) \times 10^{15}M_\odot /h$ from
the inferred member galaxies of SDSS Stripe 82 data. If we re-do the
cluster mass estimate from the measured tangential shear profile,
assuming the cluster mass definition $M_{500\rho_c}$ ($500$ times the
critical density) and the same $\Lambda$CDM cosmology \cite{Reese:2012}
used ($\Omega_m = 0.3$ and $H_0 = 70{\rm km}\ {\rm s}^{-1}\ {\rm
Mpc}^{-1}$), we find $0.59^{+0.23}_{-0.21} \times 10^{15} M_\odot$ /h,
which is consistent with the mass estimates in \cite{Reese:2012}.

In Fig.~\ref{fig:sz_m_scaling_ralation}, we compare our lensing mass
estimate for ACTJ0022 with the mass estimates of \cite{Reese:2012} in
the Compton-$y$ and cluster mass plane, also comparing with other
results for low-$z$ clusters in \cite{Marrone:2011}.  The Compton-$y$
parameter we quote is defined as
\begin{equation}
Y_{\Delta \rho_c} \equiv \frac{k_B \sigma_T}{m_e c^2}\int^{r < r_{\Delta \rho_c}}{n_e(r)T_e(r) \mathrm{d}V},
\end{equation}
where $T_e$, $m_e$, and $n_e$ are temperature, mass, and density of
elections in the hot cluster gas , and $\sigma_T$ is the cross section
of Thomson scattering. \cite{Marrone:2011} compared the $y$-parameter
derived from the SZA observations with the weak lensing masses in
\cite{Okabe:2010} for 18 X-ray luminous clusters in the redshift range
$z=[0.15,0.3]$, which are in the LoCuSS
sample\footnote{\url{http://www.sr.bham.ac.uk/locuss/}}. Then they
derived the scaling relation assuming the power-law form, denoted by the
solid line. Our weak lensing mass of ACTJ0022 seems to prefer a higher
mass for a fixed $Y_{\Delta \rho_c}$ than the scaling relation, but
again not significant due to the large error bars. The dotted and dashed
lines show the scaling relations derived in \cite{Arnaud:2010} and
\cite{Andersson:2011}, which are based on the X-ray observations for
low-$z$ clusters ($z\simlt 0.2$) and the SPT SZ clusters,
respectively. In particular, \cite{Andersson:2011} made the follow-up
Chandra and XMM-Newton observations of the 15 SPT-selected clusters,
which cover a wide range of redshifts up to $z=1$ and has the mean
redshift of $0.67$. The dot-dashed line shows the scaling relations
derived in \cite{sifon:2012}, which is based on the dynamical mass
estimates for the ACT SZ clusters ranging from $z=0.28$ to $z=1.06$ with
the mean redshift of $0.55$. Our weak lensing mass of ACTJ0022, which is
also a high-$z$ SZ cluster, seems to lie closer to the scaling relation
of \cite{Andersson:2011} and \cite{sifon:2012}, but more observations
are definitely needed to derive a more robust conclusion.

\subsubsection{$\Lambda$CDM Exclusion Curve}
\begin{figure}
\begin{center}
\includegraphics[width=9cm]{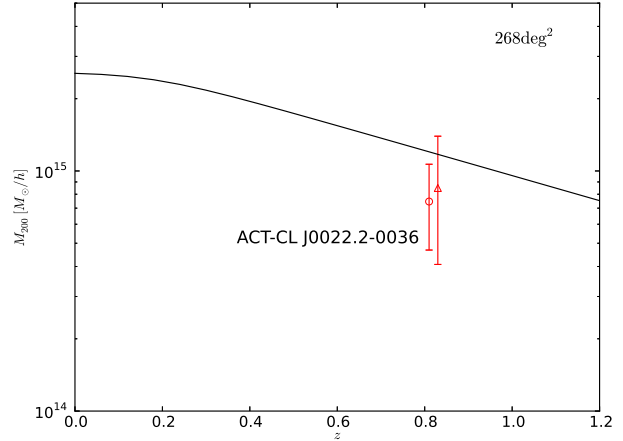}
\caption{ $\Lambda$CDM-model derived exclusion curve with 95\% C.L.  for
cluster mass of the most massive cluster in a survey area of 268 square
degrees, close to the ACT survey area, which is computed based on the
method in \protect\cite{Mortonson:2011}. If any cluster is found to have
its mass above the curve, it gives conflict with the $\Lambda$CDM model
that is consistent with other various observations.  The circle and
diamond points (slightly shifted horizontally for clarity) denote our
weak lensing mass estimates for ACTJ0022 (see Table~\ref{tab:nfwfit}),
and the error bars show 1$\sigma$ statistical error.}
\label{fig:exclusion_curve}
\end{center}
\end{figure}

ACTJ0022 is one of the most luminous SZ clusters at high redshift. The
existence of 
massive and higher redshift clusters gives a stringent
test of the $\Lambda$CDM structure formation model as well as the nature
of the primordial perturbations. Here we use the method in
\cite{Mortonson:2011} to address whether or not the existence of ACTJ0022
 is consistent with the $\Lambda$CDM prediction. To make the
test, we use the weak lensing mass estimate rather than the X-ray or
SZ-derived masses.

Fig.~\ref{fig:exclusion_curve} shows the result. The solid curve shows
the 95\% C.L. confidence level curve, computed using the code publicly
available from the
website\footnote{\url{http://background.uchicago.edu/abundance/}}\citep{Mortonson:2011};
if any cluster lying above the curve were found, it could falsify or at
least challenge the standard $\Lambda$CDM model that is constrained by
observations such as CMB, SNe, and BAO. To compute the confidence curve,
we assumed 268 square degrees for the ACT survey region overlapping with
SDSS Stripe 82. The circle and triangle symbols are the mass estimates
for Case 1 and 2 in Table~\ref{tab:nfwfit}, which are both under the
exclusion curve. Hence the existence of ACTJ0022 is consistent with the
$\Lambda$CDM model. Note that Eddington bias is not corrected in
Fig.~\ref{fig:exclusion_curve}. However, since the mass function steeply
falls with mass, the mass is reduced after the correction.  Thus,
the conclusion we made above does not change.

\section{Conclusions}
\label{sec:conclusions}

In this paper, we have used multi-band ($Br'i'z'Y$) Subaru images to
study the weak lensing signal for ACTJ0022 at $z=0.81$, which is
one of the most luminous SZ clusters identified by the ACT survey. By
using photometric redshifts derived from the multi-band data, we built a
robust catalog of background galaxies behind the high-redshift cluster,
leaving us a lower number of background galaxies, $3.2$~arcmin$^{-2}$
compared to the original density of about $20$~arcmin$^{-2}$ for all the
galaxies usable for weak lensing analysis. Nevertheless, we detected the
lensing distortion signal at $3.7\sigma$, suggesting that the
SZ-luminous ACTJ0022 is a massive cluster with virial mass $M_{200}\sim
0.8\times 10^{15}~M_\odot/h$ (see Fig.~\ref{fig:shear} and
Table~\ref{tab:nfwfit}).

While the statistical significance of this detection is not high, we
nonetheless were careful in how we did the weak lensing analysis. First,
we developed a method of using different exposure images to model the
shape of each galaxy image. In this simultaneous multi-exposure fitting
method, we can use the same model parameters for each galaxy over the
different exposures, and can use the PSFs from each exposure, which
allows us to keep the highest-resolution images and avoid a mixture of
different PSFs. Note that we did use the stacked image for object
detection.  Due to the gain in the spatial resolution, we can use
slightly smaller-size galaxies for the lensing shape measurement, by
about 10 per cent, compared to the analysis using the stacked image,
where each galaxy image is more affected by the PSF smearing effect,
especially the worst-seeing exposure. We also developed a diagnostic
method of using the star-galaxy correlation {\em residual} function, in
order to identify particular exposures that may cause systematic error
in the shape measurement (see Section~\ref{sec:residual_PSF} and
Fig.~\ref{fig:residual_correlation}). Secondly, in the PSF and galaxy
shape measurements, we included astrometric, optical distortion effect
by fitting the star and galaxy images in the celestial coordinates . The
astrometric distortion is treated as a coordinate transformation between
the CCD pixel coordinates and the celestial coordinates , not a
convolution effect as for the PSF smearing. We believe that these
methods can potentially improve our ability to accurately measure the
PSF and galaxy shapes, by minimizing systematic errors, which are
desired for upcoming wide-area weak lensing surveys such as the Subaru
HSC survey and DES.

Our method of estimating galaxy photometry for photo-$z$ was also
designed to minimize systematic error. Following the method in
\cite{Hildebrandt:2012}, we measured the 
color of every galaxy in the {\em same} physical region. To do this, we
made a PSF matching/homogenization for the stacked images of different
passbands in order to have the same PSFs in different passbands and
across different positions in each image (see Table~\ref{tab:PSFmatch}).
Then, by defining the same aperture region 
around each galaxy which is defined in the WCS, we could measure the
color of the galaxy in the same region. We tested our photo-$z$ estimate
by comparing with the spectroscopic redshifts of cluster members taken
with Gemini/GMOS, which shows a good agreement with our photo-$z$
estimates (see Fig.~\ref{fig:n_zp}). However, since we also found
photo-$z$ outlier contamination at $z_p>2$, we imposed a rather
stringent cut on the photo-$z$, $0.95<z<2.0$ to define a secure catalog
of background galaxies used for our lensing analysis. For the data
reduction/image processing, we used tools from the pipeline being
developed for the HSC survey.

Our lensing mass estimates for ACTJ0022, $M_{200} = 0.75
^{+0.32}_{-0.28} \times 10^{15}M_\odot/h$ ($M_{200} =0.85
^{+0.55}_{-0.44} \times 10^{15}M_\odot/h$) for the NFW fitting with a
free (fixed to 4.0) concentration parameter, are consistent with the
mass estimates from the SZA observations assuming hydrostatic
equilibrium and from the optical richness-WL mass scaling relation in
\cite{Reese:2012}, within the statistical measurement errors
(Fig.~\ref{fig:sz_m_scaling_ralation}). We also discussed what the mass
estimate for the high-$z$ SZ cluster ACTJ0022 may imply for the scaling
relation of the Compton-$y$ and cluster mass, comparing with the scaling
relations studied in previous works \citep{Andersson:2011, Arnaud:2010,
Marrone:2011}. The cluster observable and mass scaling relation is of
critical importance for cluster cosmology \citep{Weinberg:2012}.  Our
study is the first step towards building the SZ and WL mass scaling
relation for high-$z$ clusters, and we must increase the size of the
sample of high-redshift SZ clusters to study their WL signals.  Joint
optical and SZ experiments will be increasingly important for the
upcoming surveys, the Subaru HSC survey and the DES, which overlap the
ACT and SPT SZ surveys.

Our result demonstrates the difficulty in obtaining a high
signal-to-noise ratio WL measurements for individual high-$z$ clusters,
due to the small number density of background galaxies and photo-$z$
limitations. In addition, there are physical effects that cause
systematic issues for the WL mass estimate of individual clusters;
projection effects and aspherical mass distributions that are
unavoidable even for a perfect WL measurement. To overcome these
obstacles, we can use stacked lensing measurement or cluster-shear
correlation function method in order to boost the WL signal-to-noise
ratios and remove the systematic errors after the statistical average
\citep{Oguri:2011,Mandelbaum:2012}.  For upcoming surveys such as the
HSC, we can expect to find over a thousand massive clusters with $\ge
10^{14}M_\odot$ at $z>1$ over 1500 square degrees. Such a stacking
analysis will be powerful to obtain the average cluster mass as well as
study the scaling relations of the WL mass and cluster observables as a
function of the binned cluster observables \citep{Fang:2012}.  The
stacked lensing is based on a careful WL analysis of individual cluster
regions, but the increased $S/N$ coming from the stack makes us more
sensitive to systematic errors; thus, the methods we developed in this
paper will be useful for upcoming surveys.

\section*{Acknowledgments}

H.M. and M.T. greatly thank Gary Bernstein, Bhuvnesh Jain and 
Mike Jarvis for
many useful and constructive discussion 
on the shape measurement methods. 
H.M. acknowledges support by MEXT/JSPS Grant-in-Aid for JSPS Fellows
(DC1).  This work is supported in part by JSPS KAKENHI (Grant Number:
23340061), JSPS Core-to-Core Program ``International Research Network
for Dark Energy'', by World Premier International Research Center
Initiative (WPI Initiative), MEXT, Japan, and by the FIRST program
``Subaru Measurements of Images and Redshifts (SuMIRe)'', CSTP, Japan.

This work is based in part on data collected at Subaru Telescope, which
is operated by the National Astronomical Observatory of Japan.

Funding for SDSS-III has been provided by the Alfred P. Sloan Foundation, the Participating Institutions, the National Science Foundation, and the U.S. Department of Energy Office of Science. The SDSS-III web site is http://www.sdss3.org/.

SDSS-III is managed by the Astrophysical Research Consortium for the Participating Institutions of the SDSS-III Collaboration including the University of Arizona, the Brazilian Participation Group, Brookhaven National Laboratory, University of Cambridge, Carnegie Mellon University, University of Florida, the French Participation Group, the German Participation Group, Harvard University, the Instituto de Astrofisica de Canarias, the Michigan State/Notre Dame/JINA Participation Group, Johns Hopkins University, Lawrence Berkeley National Laboratory, Max Planck Institute for Astrophysics, Max Planck Institute for Extraterrestrial Physics, New Mexico State University, New York University, Ohio State University, Pennsylvania State University, University of Portsmouth, Princeton University, the Spanish Participation Group, University of Tokyo, University of Utah, Vanderbilt University, University of Virginia, University of Washington, and Yale University.

\appendix


\bibliography{aamnem99,ms}

\end{document}